\documentclass[
showkeys,
superscriptaddress,
preprint,
amsmath,amssymb,
aps,
prb,
floatfix,
]{revtex4-2}

\PassOptionsToPackage{backend=biber}{biblatex}
\usepackage[T1]{fontenc}
\usepackage[utf8]{inputenc}

\usepackage{siunitx}
\usepackage{amsmath} 
\usepackage{bm}

\usepackage{xcolor}
\usepackage[pdfencoding=auto]{hyperref}
\hypersetup{colorlinks=true, citecolor=blue, urlcolor=blue, linkcolor=black}

\usepackage{xr}
\makeatletter
\newcommand*{\addFileDependency}[1]{
  \typeout{(#1)}
  \@addtofilelist{#1}
  \IfFileExists{#1}{}{\typeout{No file #1.}}
}
\makeatother

\newcommand*{\myexternaldocument}[2][]{%
    \externaldocument[#1]{#2}%
    \addFileDependency{#2.tex}%
    \addFileDependency{#2.aux}%
}
\myexternaldocument[supp-]{supplementary}

\usepackage{graphicx} 
\graphicspath{ {DGraphics/} }

\usepackage[english]{babel}  

\usepackage[caption=false]{subfig}
\usepackage[compatibility=false]{caption}  
\usepackage{wrapfig}
\usepackage[export]{adjustbox}
\usepackage{relsize}
\usepackage{multirow}
\usepackage{dcolumn}
\usepackage{array}

\usepackage[normalem]{ulem}
\usepackage{booktabs}
\usepackage{makecell}

\renewcommand{\figurename}{Figure}

\newcommand{\figref}[2][\figurename~]{#1\ref{#2}}
\newcommand{\figrefs}[3][\figurename~]{#1\ref{#2},\subref*{#3}}
\newcommand{\Figrefs}[3][\figurename~]{#1\ref{#2},\subref*{#3}}
\newcommand{\figrefspan}[3][\figurename~]{#1\ref{#2}--\subref*{#3}}

\newcommand{\suppsecref}[2][Section~]{\hyperref[supp-#2]{#1\ref*{supp-#2}}}
\renewcommand{\eqref}[2][Equation~]{#1(\ref{#2})}

\DeclareSIUnit\angstrom{\protect\text{Å}}
\DeclareSIUnit{\rpm}{rpm}

\newcolumntype{P}[1]{>{\centering\arraybackslash}p{#1}}

\definecolor{amethyst}{rgb}{0.6, 0.4, 0.8}

\begin{document}

\title{Compact Variational Neural Networks for Spectral Inference from a Single Nonlinear 2D Perovskite Photodetector}

\author{Karl Jonas Riisnaes}
\altaffiliation{These authors contributed equally to this work.}
\affiliation{Centre for Graphene Science, Department of Physics and Astronomy, University of Exeter, Exeter EX4 4QL, United Kingdom}

\author{Ned Thaddeus Taylor}
\altaffiliation{These authors contributed equally to this work.}
\affiliation{Centre for Graphene Science, Department of Physics and Astronomy, University of Exeter, Exeter EX4 4QL, United Kingdom}

\author{Hoi Tung Lam}
\affiliation{Centre for Graphene Science, Department of Physics and Astronomy, University of Exeter, Exeter EX4 4QL, United Kingdom}

\author{Rosanna Mastria}
\affiliation{CNR NANOTEC, Institute of Nanotechnology, Via Monteroni, 73100 Lecce, Italy}

\author{Francesco Saverio Difeo}
\affiliation{CNR NANOTEC, Institute of Nanotechnology, Via Monteroni, 73100 Lecce, Italy}
\affiliation{Department of Mathematics and Physics “Ennio de Giorgi”, University of Salento, via Monteroni, 73100, Lecce, Italy}

\author{Monica Felicia Craciun}
\affiliation{Centre for Graphene Science, Department of Engineering, University of Exeter, Exeter EX4 4QL, United Kingdom}

\author{Saverio Russo}
\email{s.russo@exeter.ac.uk}
\affiliation{Centre for Graphene Science, Department of Physics and Astronomy, University of Exeter, Exeter EX4 4QL, United Kingdom}

\date{\today}

\begin{abstract}
Spectroscopy conventionally separates optical frequencies before detection, imposing persistent constraints on footprint, complexity and scalability.
Here we establish an alternative paradigm in which the nonlinear optoelectronic dynamics of a single two-dimensional perovskite photodetector physically encode the incident optical field and machine learning performs the inverse spectral reconstruction.
Using a planar fluorinated phenethylammonium lead iodide (F-PEAI) photodetector, we exploit wavelength- and irradiance-dependent current-voltage signatures arising from the coupled effects of photocarrier generation, trapping, interfacial transport and field-dependent carrier dynamics.
A compact variational encoder-decoder preserves the functional and history-dependent structure of these responses by independently projecting forward and reverse voltage sweeps onto a truncated Legendre-polynomial basis before mapping them through a probabilistic latent representation to continuous spectral parameters.
Trained on fewer than 400 experimental voltage sweeps, the model generalises to excitation wavelengths excluded from training, reconstructing wavelength with $R^2=0.958$ and a mean absolute error of 8.1~\si{\nano\metre}, while recovering log-normalised irradiance with $R^2=0.987$.
Voltage-resolved analysis further reveals that wavelength and irradiance are encoded differently across the nonlinear device response, with distinct bias regions carrying complementary optical information.
These results establish nonlinear material and interface dynamics as a computational resource for spectroscopy and point towards hardware-algorithm co-design in which materials, interfaces and inference architectures are engineered jointly to maximise information content, enabling compact spectroscopic systems without dispersive optics or detector arrays.

\end{abstract}

\keywords{machine learning, photodetector, nonlinear, photocurrent, variational autoencoder, encoder-decoder, self-supervised learning, physics-inspired, neural network, uncertainty quantification}

\maketitle

\section{Introduction}
Spectroscopy underpins modern science and technology by revealing the chemical composition and physical state of matter through its interaction with light.
From benchtop instruments to integrated photonic devices, a large class of conventional spectrometers operate according to the same fundamental principle: spectral information is mapped into space by dispersive or wavelength-selective optical elements and subsequently recorded by arrays of linear photodetectors~\cite{Zhang2025ReconstructiveSpectrometersHardware,Yang2021MiniaturizationOpticalSpectrometers}.
While continual miniaturisation has yielded increasingly compact implementations, this architecture remains fundamentally constrained by the physics of optical dispersion, whose resolving power scales with the dimensions of the dispersive element.
As a consequence, conventional spectrometers face a persistent trade-off between footprint, complexity, and spectral performance~\cite{Xue2024AdvancedsMiniaturizedComputational}, limiting their deployment in emerging applications requiring ultracompact, low-power, and scalable sensing platforms.
Computational spectroscopy offers an alternative strategy in which the incident spectrum is encoded into a calibrated optical or electrical response and subsequently recovered through numerical inference~\cite{Xue2024AdvancedsMiniaturizedComputational,Wang20252DComputationalPhotodetectors, wangMiniaturizedSpectrometerBased2024}.
Its performance therefore depends on both the ability of the sensing hardware to generate distinguishable responses to different optical inputs and the capacity of the reconstruction algorithm to invert this encoding accurately in the presence of experimental noise and limited training data.

A particularly promising route to computational spectroscopy is to exploit the nonlinear photoresponse of the detector itself as the spectral-encoding mechanism~\cite{Li2025NonlinearMemristiveComputational}.
In photoactive semiconductors, photocarrier generation, trapping and detrapping, contact-mediated transport, electric-field redistribution and illumination or bias history can collectively produce complex current-voltage responses whose shape depends on both photon energy and irradiance~\cite{Iddo2017,Townsend_2018,Intonti}.
Phenomena such as nonlinearity, hysteresis and contact asymmetry, which were conventionally regarded as detrimental, can become information-bearing degrees of freedom.
Two-dimensional (2D) hybrid perovskites are particularly attractive for this approach because strong excitonic light-matter interactions coexist with trap-mediated transport, field-dependent carrier dynamics and pronounced sensitivity to metal-semiconductor interfaces~\cite{Blancon2020,Stoumpos,Durante_AdvFuncMater2025,Mastria2026WaferScaleRoom}, while their compatibility with high-resolution electron-beam lithography enables reproducibly engineered planar device geometries~\cite{Mastria2024RealTimeHighly}. In these materials, wavelength modulates absorption and the microscopic pathways of carrier generation and transport, whereas irradiance changes photocarrier density, trap occupation and the resulting transport regime~\cite{Leontis2026DecodingTrapStates}. The resulting bias-dependent electrical trajectory can therefore encode substantially more optical information than a single linear-intensity measurement~\cite{DeSanctisLinearDynamicRange,Jones}, but recovering the incident optical state from this multidimensional response constitutes a nonlinear inverse problem naturally suited to data-driven inference.

Indeed, recent studies have established the feasibility of computational spectroscopy based on nonlinear photodetectors by combining machine learning with complex electrical responses to recover spectral information directly from a single sensing element~\cite{Darweesh2024NonlinearSelfCalibrated,Chen2025MultiTaskNeural,Oter2026StackingEnsembleMachine}.
More broadly, latent-space learning has proven highly effective for modelling complex current-voltage characteristics in optoelectronic devices, demonstrating the potential of physics-aware machine-learning representations for nonlinear transport~\cite{Zbinden2025AutoencoderParameterEstimation}.
However, existing implementations are generally tailored to particular devices and measurement grids and often treat the sampled electrical response as a generic high-dimensional vector or predict discrete wavelength classes or spectral bins.
Wang et~al.~\cite{Wang20252DComputationalPhotodetectors} emphasised the potential of 2D materials for computational spectroscopy and identified detection performance, perception functionality, and chip-level integration as key priorities for future development~\cite{Wang20252DComputationalPhotodetectors}.
A remaining challenge is to develop compact architectures that can simultaneously extract and interpret multidimensional optical parameters (i.e. intensity, wavelength, and beyond) from the complex nonlinear relationship between optical inputs and electrical outputs~\cite{Wang20252DComputationalPhotodetectors}, whilst providing meaningful measures of prediction confidence.

Here we present a computational spectroscopy platform in which the nonlinear optoelectronic response of a single 2D fluorinated phenethylammonium lead iodide (F-PEAI) photodetector acts as a physical encoder of the incident optical field.
2D F-PEAI combines strong light-matter interaction and room-temperature excitonic response with trap-mediated transport, field-dependent carrier dynamics and pronounced metal-semiconductor interface effects, generating wavelength- and irradiance-dependent current-voltage signatures that extend beyond a simple linear responsivity~\cite{Leontis2026DecodingTrapStates,Mastria2024RealTimeHighly,Riisnaes20242DHybridPerovskite}.
To decode this response, we develop a compact variational encoder-decoder that preserves both the functional form and history dependence of the measurement: forward and reverse voltage sweeps of the corrected photocurrent and dark current are independently projected onto a truncated Legendre-polynomial basis before being mapped through a probabilistic latent representation to continuous spectral parameters.
This structure-aware formulation is not tied to a fixed number of voltage samples and avoids discretising the output onto a predefined wavelength grid, while latent-space sampling provides an internal measure of prediction sensitivity.
The model infers continuous wavelength and irradiance simultaneously. 
Trained on fewer than 400 experimental voltage sweeps, the model generalises to excitation wavelengths excluded from training, reconstructing wavelength with $R^2=0.958$ and a mean absolute error of 8.1~\si{\nano\metre}, while predicting log-normalised irradiance with $R^2=0.987$.
Remarkably, irradiance remains recoverable even after the absolute current magnitude of each sweep is normalised away, showing that optical information is encoded not simply in signal amplitude but in the multidimensional structure of the detector response.
These results establish nonlinear material and interface dynamics as a computational resource for spectroscopy, enabling spectral encoding and detection to be collapsed into a single nanoscale-compatible optoelectronic element without dispersive optics or detector arrays.

\begin{figure}
    \centering%
    \noindent%
    \subfloat[\vspace{-0.5em}]{\includegraphics[width=0.315\linewidth]{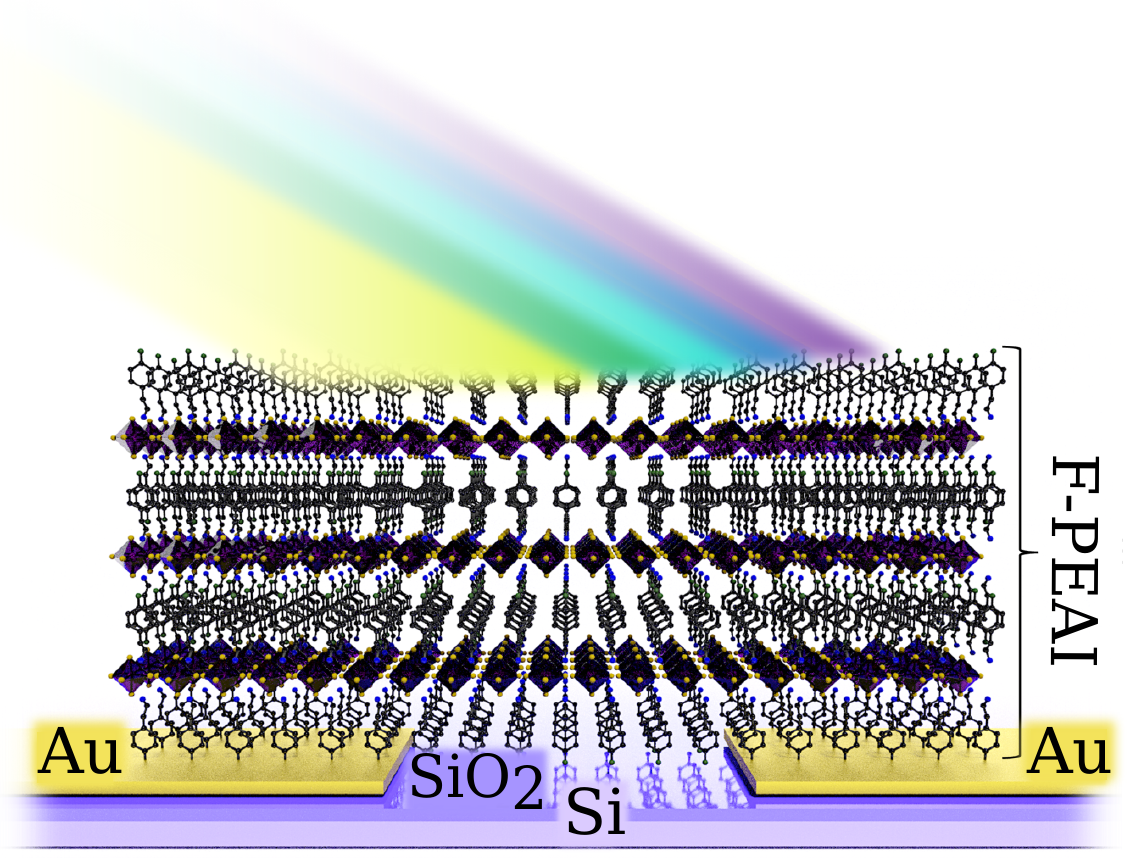}\label{fig:device:photodetector}}%
    \hspace{0.5em}%
    \subfloat[\vspace{-0.5em}]{\includegraphics[width=0.315\linewidth]{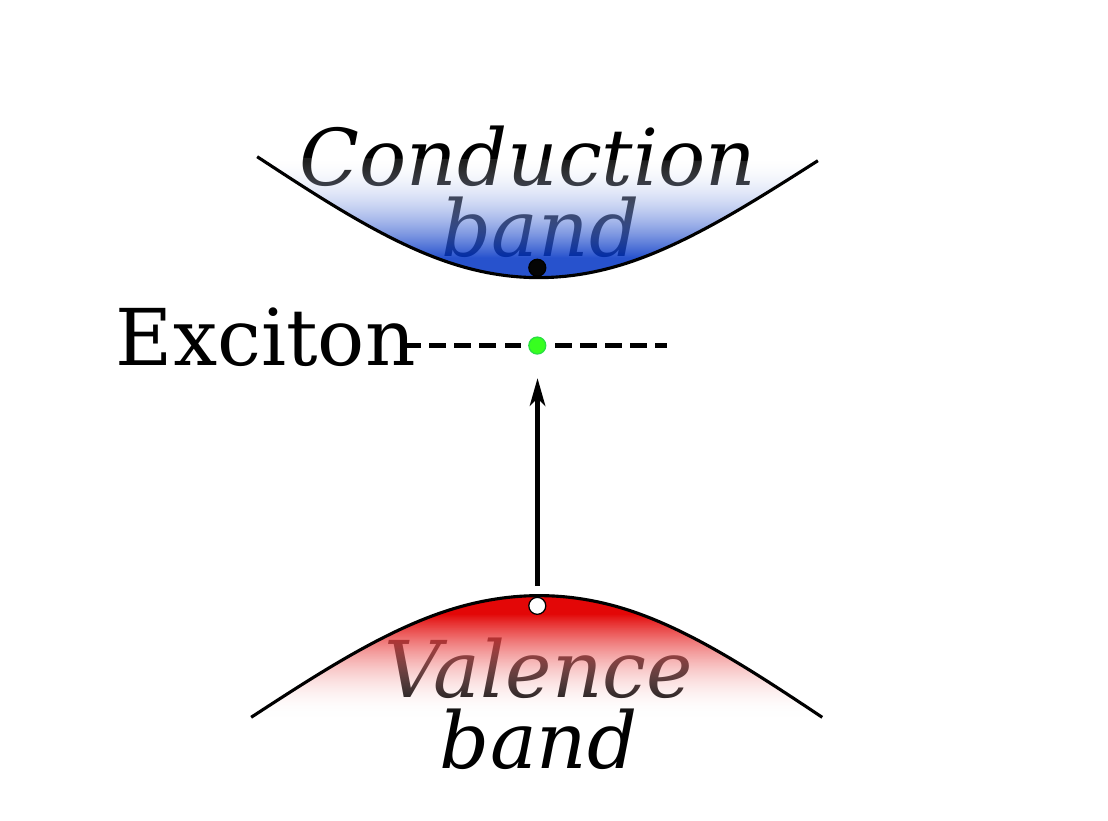}\label{fig:device:exciton}}%
    \hspace{-0.5em}%
    \subfloat[\vspace{-0.5em}]{\includegraphics[width=0.315\linewidth]{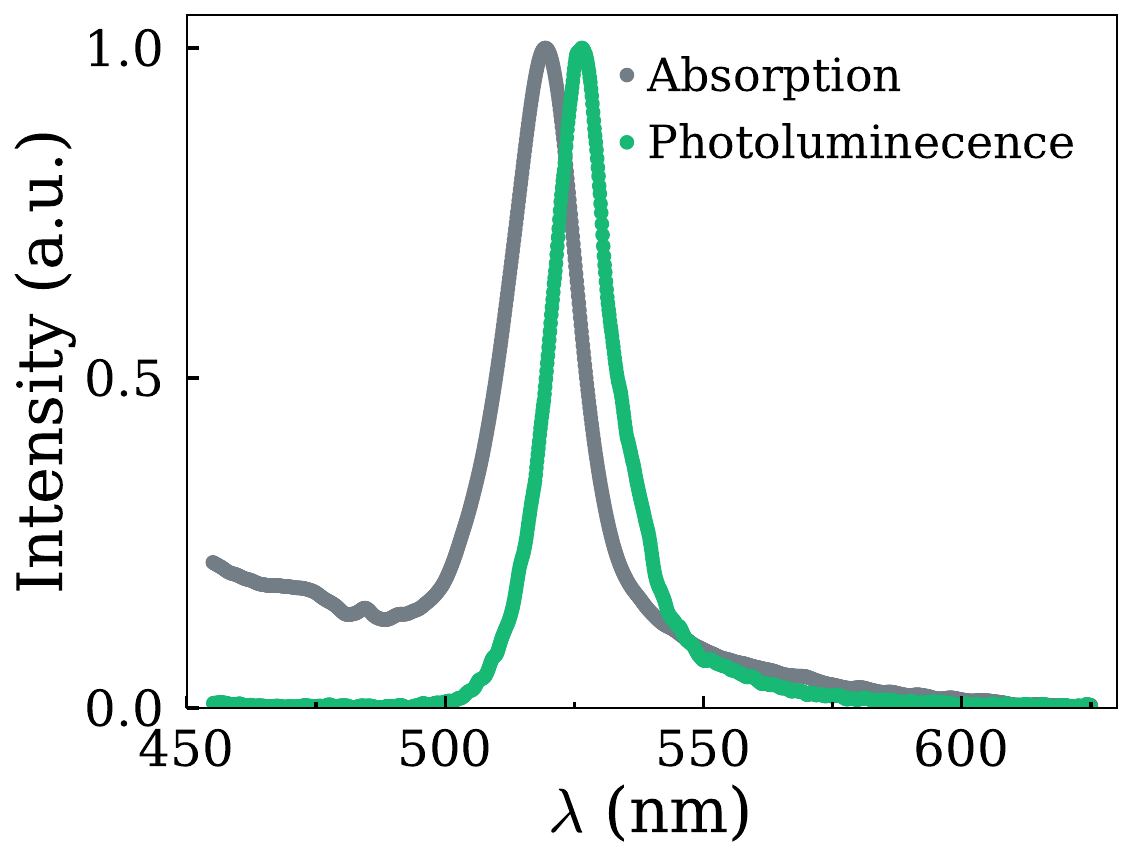}\label{fig:device:photon_counts}}%
    \vspace{-1em}\hspace{-0.5em}%
    \subfloat[\vspace{-0.5em}]{\includegraphics[width=0.315\linewidth]{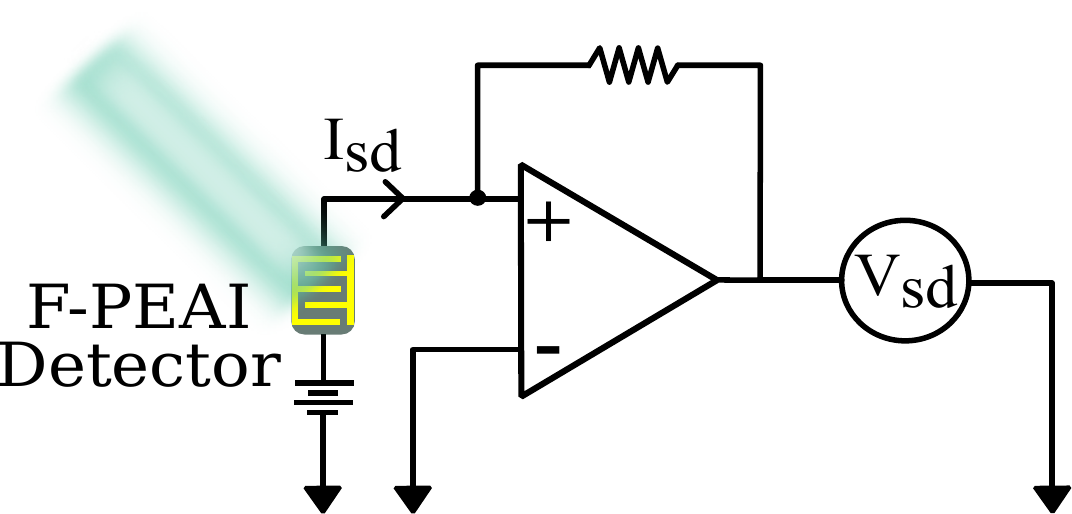}\label{fig:device:circuit}}%
    \hspace{0.5em}%
    \subfloat[\vspace{-0.5em}]{\includegraphics[width=0.315\linewidth]{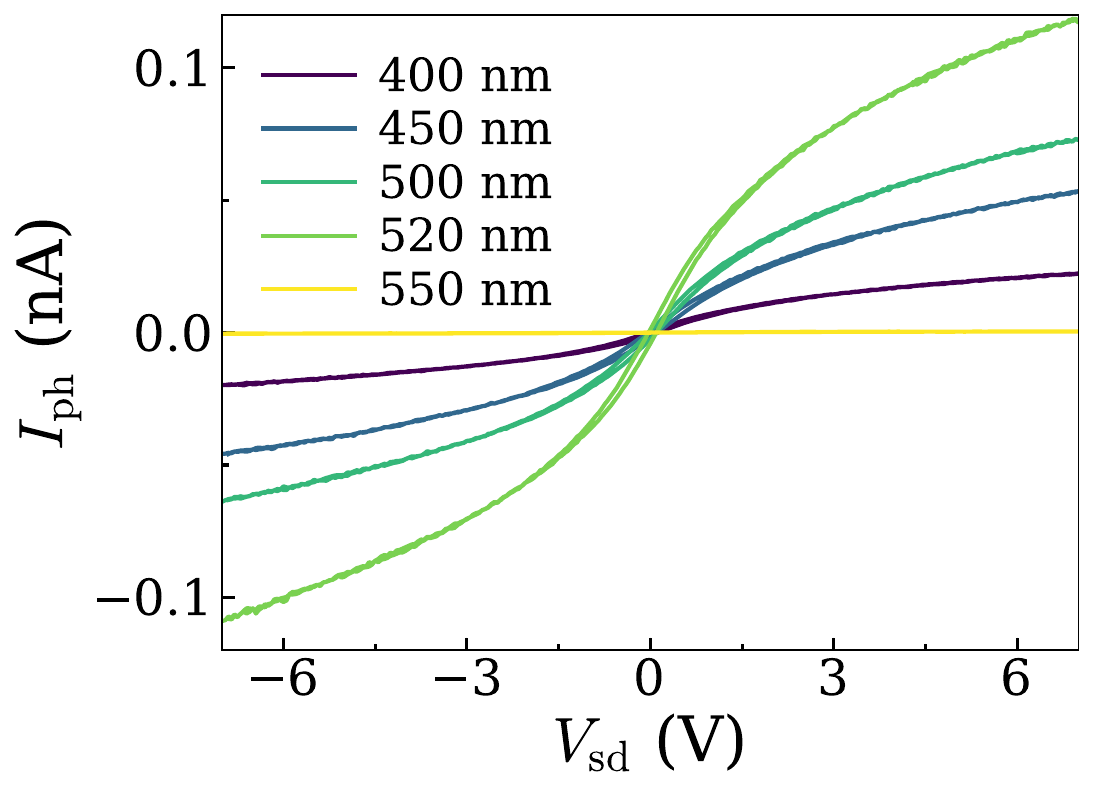  }\label{fig:device:iv_curves}}%
    \hspace{0.5em}%
    \subfloat[\vspace{-0.5em}]{\includegraphics[width=0.315\linewidth]{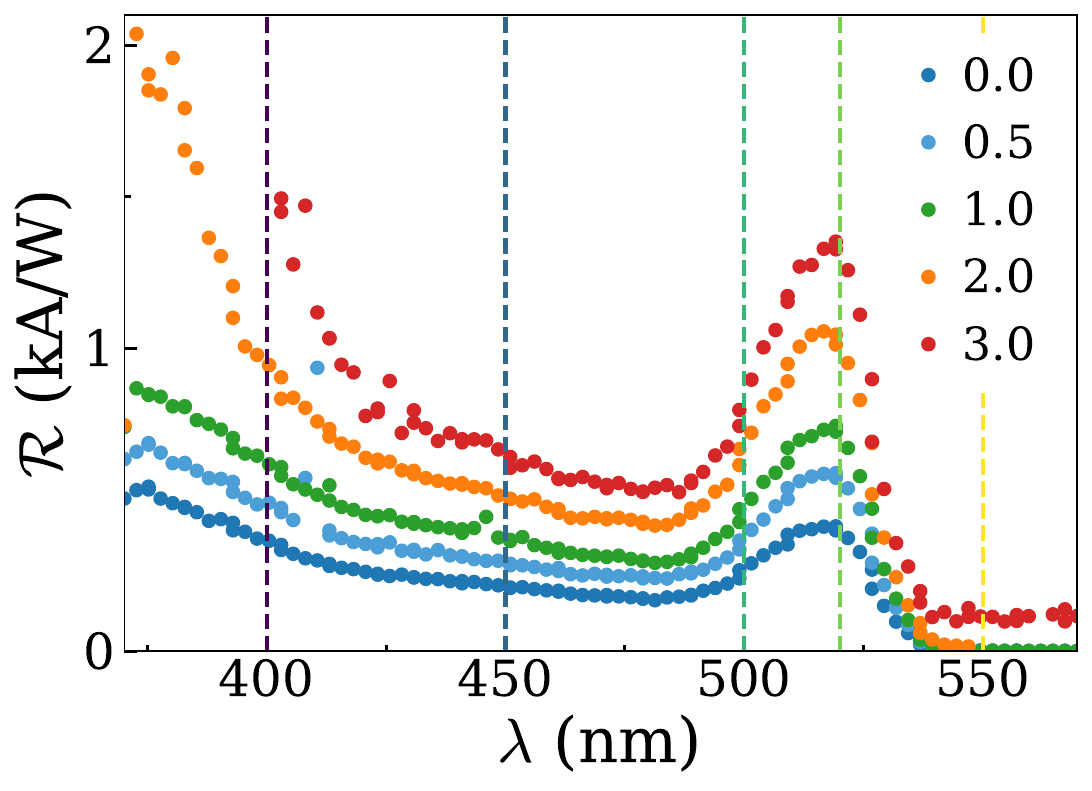}\label{fig:device:photoresponsivity}}%
    \caption{%
        \textbf{Optoelectronic response of the 2D F-PEAI photodetector.}
        \protect\subref{fig:device:photodetector}~Schematic of the planar F-PEAI device under optical excitation.
        \protect\subref{fig:device:exciton}~Schematic representation of the excitonic transition in F-PEAI.
        \protect\subref{fig:device:photon_counts}~Optical absorption and photoluminescence spectra of an F-PEAI crystal.
        \protect\subref{fig:device:circuit}~Electrical circuit used for bias-dependent photocurrent measurements.
        \protect\subref{fig:device:iv_curves}~Representative photocurrent-voltage characteristics under selected monochromatic excitation wavelengths at optical-density filter (OD) 1.0 (%
        $\lambda=400$~\si{\nano\metre} and $\mathcal{I}=4.6$~\si{\micro\watt\per\centi\metre\squared},
        $\lambda=450$~\si{\nano\metre} and $\mathcal{I}=17.8$~\si{\micro\watt\per\centi\metre\squared},
        $\lambda=500$~\si{\nano\metre} and $\mathcal{I}=20.9$~\si{\micro\watt\per\centi\metre\squared}$,
        \lambda=520$~\si{\nano\metre} and $\mathcal{I}=21.2$~\si{\micro\watt\per\centi\metre\squared},
        $\lambda=550$~\si{\nano\metre} and $\mathcal{I}=23.3$~\si{\micro\watt\per\centi\metre\squared}%
        ).
        \protect\subref{fig:device:photoresponsivity}~Absolute spectral responsivity ($\mathcal{R}$) measured at $V_{\mathrm{sd}}=-7$ V for the investigated optical-density filters; dashed lines indicate the wavelengths shown in \protect\subref{fig:device:iv_curves}.%
    }
    \label{fig:device}
\end{figure}

Among the wide range of layered halide perovskites, two-dimensional fluorinated phenethylammonium lead iodide is a particularly attractive since the fluorination of the phenethylammonium spacer confers a distinctive resilience to ambient handling and resist- and solvent-based processing~\cite{Smith_stability}, opening access to high-resolution fabrication approaches, including electron-beam lithography, that remain challenging for many metal-halide perovskites~\cite{Mastria2024RealTimeHighly,Riisnaes20242DHybridPerovskite}.
At the same time, the reduced dimensionality of the inorganic layers and strong dielectric contrast with the organic spacers produce pronounced quantum confinement~\cite{Cheng2018}, stabilising excitonic states at room temperature and concentrating optical absorption within the inorganic quantum wells~\cite{Blancon2018,Saparov2016OrganicInorganicPerovskites,Straus2018ElectronsExcitonsPhonons,Cinquino2021ManagingGrowthDimensionality,Gao2019RuddlesdenPopperPerovskites}. Consequently, F-PEAI exhibits a direct bandgap of approximately 2.61~\si{\electronvolt} ($\sim$475~\si{\nano\metre}), strong absorption across the ultraviolet and blue spectral regions, and pronounced room-temperature excitonic emission centred near 523~\si{\nano\metre} (\figrefs{fig:device:exciton}{fig:device:photon_counts}). 

For the present study, mechanically exfoliated F-PEAI flakes were transferred onto pre-patterned interdigitated Au electrodes with a total width of 16.58~\si{\milli\metre} and a channel length of 3~\si{\micro\metre} (\figref{fig:device}; see Methods and Supporting Information
Section~S1%
).
In this well-defined planar geometry, the short channel and extended contact width of the interdigitated architecture enhance the measurable photocurrent.
Bias-dependent photocurrent signatures were acquired under monochromatic illumination spanning 370--570~\si{\nano\metre}, with a linewidth of 8~\si{\nano\metre} and incident optical powers ranging from $\approx 1\times10^{-9}$ to $2.1\times10^{-5}$~\si{\watt}.
The photocurrent was collected using a current amplifier connected in series to the drain of the device (see \figref{fig:device:circuit} and Methods).
These bias-dependent dark-current and photocurrent sweeps constitute the electrical signatures used for subsequent machine-learning-based spectral reconstruction.

\section{Results}

\subsection{Photocurrent characterisation}

\Figrefs{fig:device:iv_curves}{fig:device:photoresponsivity} show representative bias-dependent photocurrent measurements and the spectral responsivity of the F-PEAI photodetector evaluated at $V_{\mathrm{sd}}=-7$~\si{\volt} (see Methods), respectively. The spectral photoresponsivity correlates well to the absorption spectrum, featuring a pronounced peak at the exciton wavelength (i.e. 516~\si{\nano\metre}), and an increase in absorption when the photon energy supports band-to-band charge excitations. Crucially, the photoresponse of 2D F-PEAI is strongly non-linear, bias- and wavelength-dependent. 
Previous studies of 2D F-PEAI photodetectors have shown that this nonlinear bias dependence arises from the interplay of charge trapping, interfacial energy barriers and field-dependent carrier transport~\cite{Leontis2026DecodingTrapStates,Mastria2024RealTimeHighly}.
The photocurrent therefore cannot be described by a single linear transfer function, but instead reflects the coupled evolution of several photoinduced processes.
A priori, this multidimensional response could be a function of the photogeneration rate $G(\lambda,\mathcal{I})$, source-drain bias $V_{\mathrm{sd}}$, the evolving electronic state of the device ($t$) during and after illumination or bias application, together with the trap-state distribution $D_t(E)$, the source- and drain-contact barrier landscape $\Phi_B^{s,d}$, and possible slowly varying ionic charge $\rho_{\mathrm{ion}}$: $I_{\mathrm{ph}}=F\left[G(\lambda,\mathcal{I}),V_{\mathrm{sd}},t;D_t(E),\Phi_B^{s,d},\rho_{\mathrm{ion}}\right]$, where $\mathcal{I}$ is the incident irradiance.
Although the measurement temperature and voltage-sweep protocol are fixed here, the internal state of the detector can evolve during a sweep through trap filling and emptying, persistent photoconductivity, carrier relaxation and possible ionic redistribution.
Wavelength and irradiance therefore influence the electrical response not only through the number of photogenerated carriers, but also through the microscopic pathways governing their generation, relaxation, trapping, transport and extraction.
Variations in absorption, exciton dissociation, trap occupation, carrier lifetime and illumination-induced modulation of the contact barriers can consequently generate distinct wavelength- and irradiance-dependent current-voltage signatures~\cite{Simbula2023,Zhou,Mehew,Iddo2017,Townsend_2018}.
This intrinsic coupling transforms the photodetector from a simple intensity-to-current transducer into a complex physical encoder of the incident optical field.

\subsection{Model}

\begin{figure}
    \centering
    \subfloat[\vspace{-2.25em}]{\includegraphics[scale=0.25]{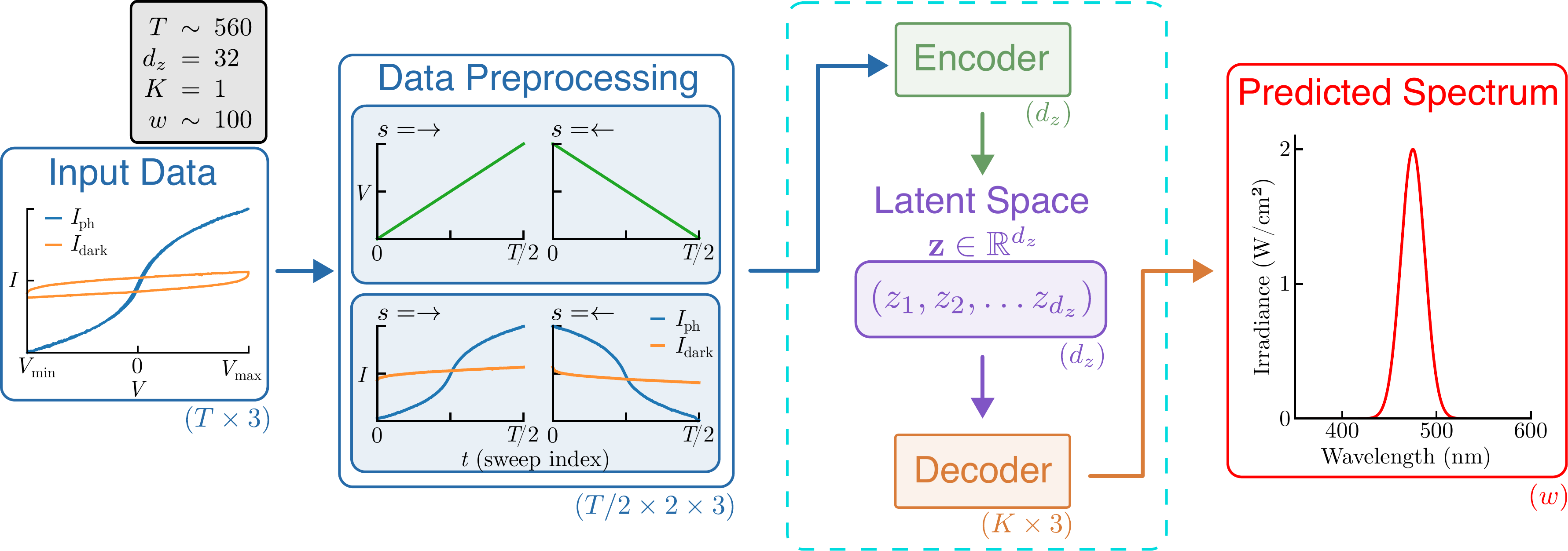}\label{fig:architecture:model}}\vspace{-1.25em}\\%
    \subfloat[\vspace{-0.65em}]{\includegraphics[scale=0.25]{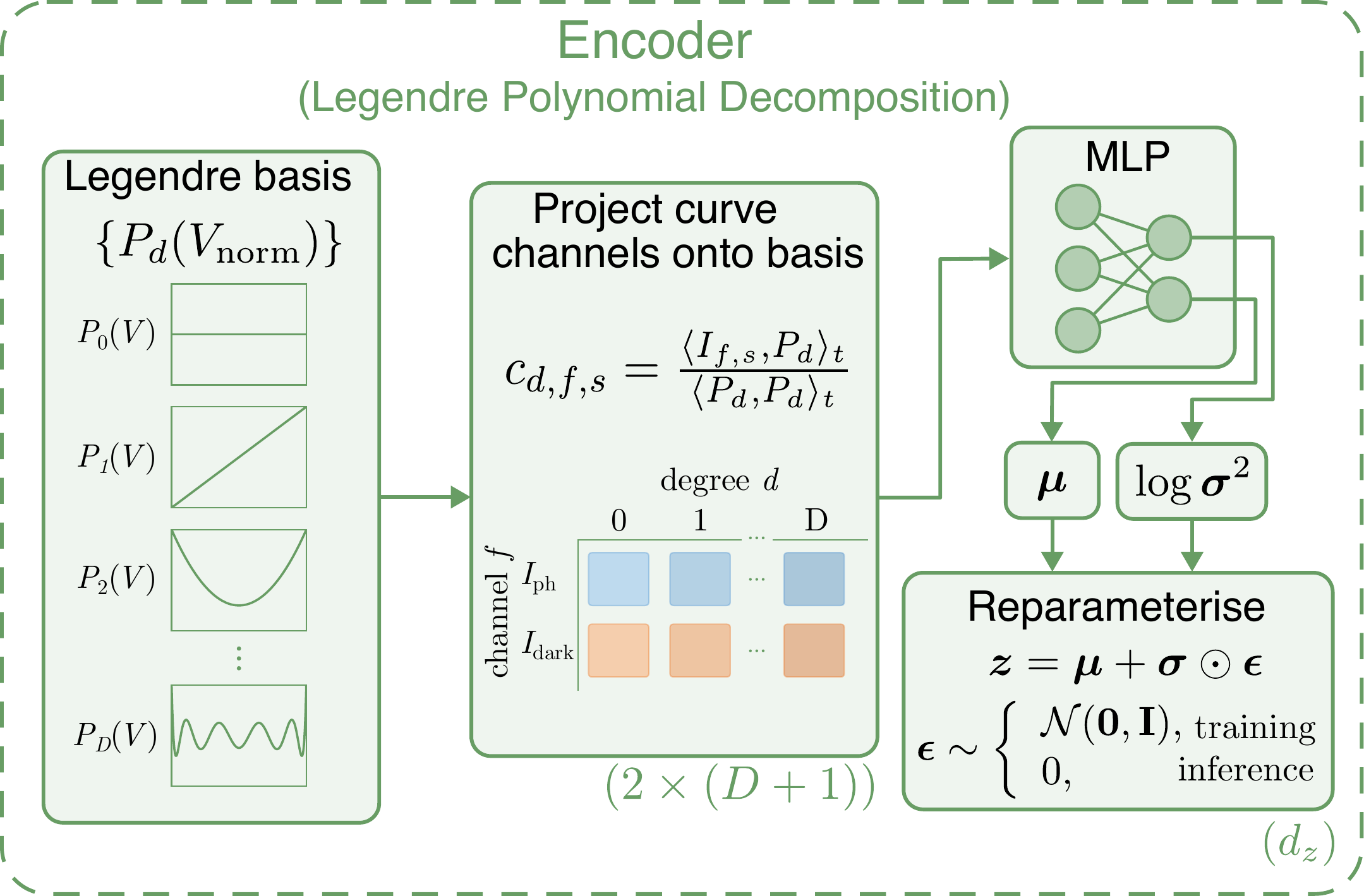}\label{fig:architecture:encoder}}\hspace{1em}%
    \subfloat[\vspace{-0.65em}]{\includegraphics[scale=0.25]{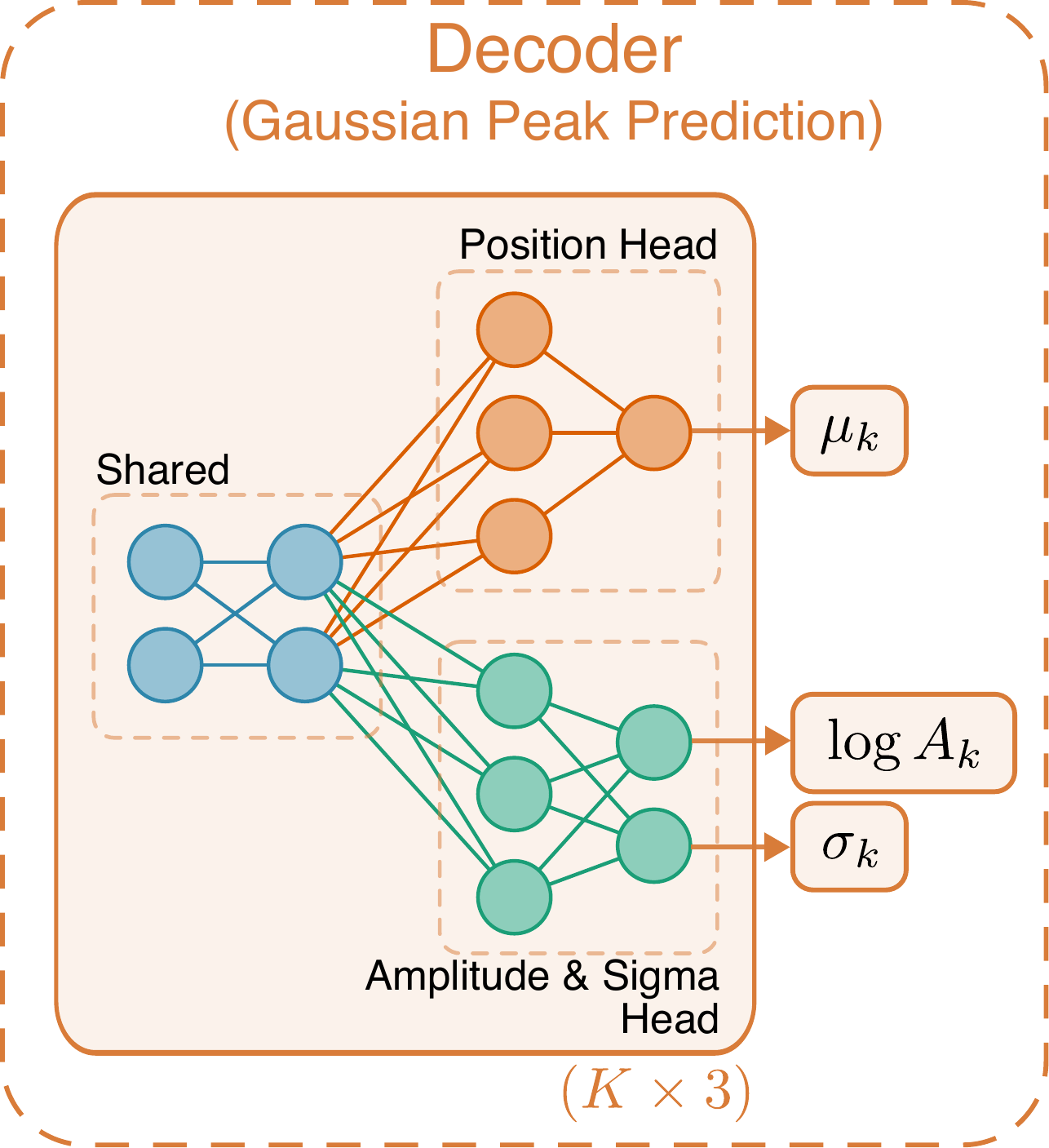}\label{fig:architecture:decoder}}
    \caption{%
        \textbf{Structure-aware variational encoder-decoder for spectral reconstruction.}
        \protect\subref{fig:architecture:model}~Overall workflow from the measured voltage, corrected photocurrent and dark-current traces to the reconstructed optical spectrum.
        Forward and reverse bias sweeps are preprocessed separately before encoding into a probabilistic latent representation and decoding into spectral parameters.
        \protect\subref{fig:architecture:encoder}~Encoder architecture: each current channel is projected onto a Legendre-polynomial basis up to degree $D$, and the resulting coefficients are mapped to the latent mean $\boldsymbol{\mu}$ and variance $\boldsymbol{\sigma}^2$.
        \protect\subref{fig:architecture:decoder}~Gaussian decoder mapping the latent representation onto the centre wavelength, amplitude and width of $K$ spectral components.%
    }
    \label{fig:architecture}
\end{figure}

Recovering the incident optical excitation from the measured current-voltage characteristics constitutes a nonlinear inverse problem for which no tractable analytical inversion is available.
We therefore develop a variational autoencoder (VAE)-inspired encoder-decoder that maps the complete bias-dependent dark-current and photocurrent response onto continuous parameters describing the incident optical field, see \figref{fig:architecture}.
The architecture adopts the Gaussian latent representation and reparameterised sampling central to VAEs, with Kullback-Leibler divergence~\cite{Kingma2013AutoEncodingVariational,Rezende2014StochasticBackpropagationApproximate} regularisation applied during self-supervised pretraining, but repurposes this framework for supervised cross-domain inference rather than reconstruction of the electrical input.
An analytical Legendre-polynomial front end transforms the measured response into a compact representation, treating the forward and reverse voltage sweeps independently to preserve their global shape, nonlinearity, and hysteresis.
This representation is encoded into the probabilistic latent space and decoded directly into a parameterised Gaussian spectrum, from which the centre wavelength and irradiance are reconstructed.
The resulting physics-informed, structured representation substantially reduces the dimensionality of both the input and output spaces, while avoiding dependence on a fixed number of sampled voltage points or a predefined wavelength grid.

The encoder is designed to retain the physical information distributed across the complete voltage sweep while removing its dependence on the discrete measurement grid.
As illustrated in \figrefs{fig:architecture:model}{fig:architecture:encoder}, the corrected photocurrent and dark-current traces are separated into forward and reverse bias branches, preserving the directional asymmetry and hysteresis associated with the evolving internal state of the photodetector.
For each current channel (dark- and photocurrent) and sweep direction, the voltage axis is mapped onto the dimensionless interval $[-1,1]$ and the corresponding current trace is projected independently onto a Legendre-polynomial basis truncated at a maximum degree $D=8$; see Supporting Information
Section~S2%
.
The resulting coefficients provide a compact representation of the dominant offset, slope, curvature and higher-order nonlinear structure of each branch.
Forward and reverse coefficient sets are retained independently rather than collapsed into a single-valued current-voltage characteristic, allowing both instantaneous bias dependence and history-dependent dynamics to contribute to the spectral inference.
Truncation at $D=8$ acts simultaneously as a dimensionality-reduction and regularisation step, suppressing point-to-point experimental noise while preserving the nonlinear and hysteretic features that encode the incident optical field.
Although fixed at $\pm 7$~\si{\volt} for the present dataset, the minimum and maximum sweep voltages are retained as input parameters to preserve the physical voltage scale after normalisation and to allow the same representation to accommodate measurements acquired over different voltage spans. The influence of the principal architectural choices is examined in Supporting Information 
Section~S7.5%
.

The resulting feature vector is mapped by the neural-network encoder onto a $32$-dimensional variational latent representation, parameterised by a learned mean $\mu$ and variance $\sigma^2$.
During training, latent states are sampled through the reparameterisation trick~\cite{Kingma2013AutoEncodingVariational,Rezende2014StochasticBackpropagationApproximate} $z=\mu+\sigma\odot\epsilon$, with $\epsilon\sim\mathcal{N}(0,I)$, whereas nominal inference would use the deterministic latent mean $z=\mu$.
The decoder subsequently transforms this latent representation into the physical parameters of a continuous Gaussian description of the incident spectrum rather than reconstructing intensities on a predefined wavelength grid.
For each spectral component $k$, the decoder predicts its centre wavelength $\mu_k$, amplitude $A_k$ and width $\sigma_k$, with the amplitude constrained in logarithmic space to the physically accessible irradiance range.
In the present monochromatic implementation, $K=1$ and the experimentally varying quantities of interest are therefore the centre wavelength and irradiance; the spectral linewidth is fixed across the dataset.
Repeated sampling from the learned latent distribution generates an ensemble of reconstructed spectra, allowing the spread of the predictions to provide a model-internal measure of sensitivity to the encoded detector state.

Training proceeds in two stages. The encoder is first pretrained using a self-supervised SimSiam objective~\cite{Chen2021ExploringSimpleSiamese}, together with variational regularisation of the latent representation, before the complete encoder-decoder is optimised using supervised learning on the physical parameters of the target spectrum. The supervised objective acts directly on Gaussian peak position, amplitude and width, while suppressing superfluous spectral components through a differentiable peak-assignment scheme. This formulation avoids point-by-point comparison on a predefined wavelength grid and naturally extends to spectra containing multiple components. Full details of the self-supervised pretraining and loss formulation are provided in Supporting Information
Sections~S3~and~S4%
, respectively.

Despite encoding the full nonlinear and history-dependent electrical response, the architecture remains compact, containing only 14,799 trainable parameters. Its physics-inspired parameterisation is central to this efficiency: Legendre projection maps voltage sweeps containing different numbers of sampled points onto a fixed-dimensional representation, while the Gaussian decoder describes the optical output through continuous spectral parameters rather than wavelength bins. The model is therefore not intrinsically tied to either the number of acquired voltage points or the spectral grid used to evaluate the reconstructed output. Although the present model is trained and evaluated exclusively on single-peak spectra ($K=1$), the decoder is formulated for an arbitrary number of Gaussian components $K$. Increasing the capacity from $K=1$ to $K=6$, for example, raises the parameter count only from 14,799 to approximately 15,279, while the training objective suppresses superfluous components when the available model capacity exceeds the spectral complexity of the input. Multi-peak reconstruction is therefore an intrinsic capability of the architecture, although its predictive performance for complex spectra remains to be experimentally established.

A further advantage of the variational formulation is that uncertainty in the encoded detector response can be propagated directly into the reconstructed spectrum, see Methods.
Although a deterministic prediction can be obtained from the latent mean $z=\mu$, the learned posterior $q_\phi(z|x)=\mathcal{N}(\mu,\mathrm{diag}(\sigma^2))$ can instead be sampled repeatedly and each latent state propagated through the same Gaussian decoder.
In the ensemble inference used here, 100 such realisations generate a distribution of reconstructed wavelengths and irradiances: the ensemble mean provides the spectral estimate, while the spread of the predictions quantifies its sensitivity to variations permitted by the learned latent representation.
This model-internal measure is distinct from the absolute reconstruction error, but provides a complementary indication of prediction stability and sensitivity to the learned latent representation.
The probabilistic architecture therefore yields not only an inverse mapping from nonlinear electrical responses to spectral parameters, but an accompanying measure of the robustness of that inference.

\subsection{Predictive Capabilities}

\begin{figure}
    \centering%
    \subfloat[\vspace{-0.75em}]{\includegraphics[scale=0.3775]{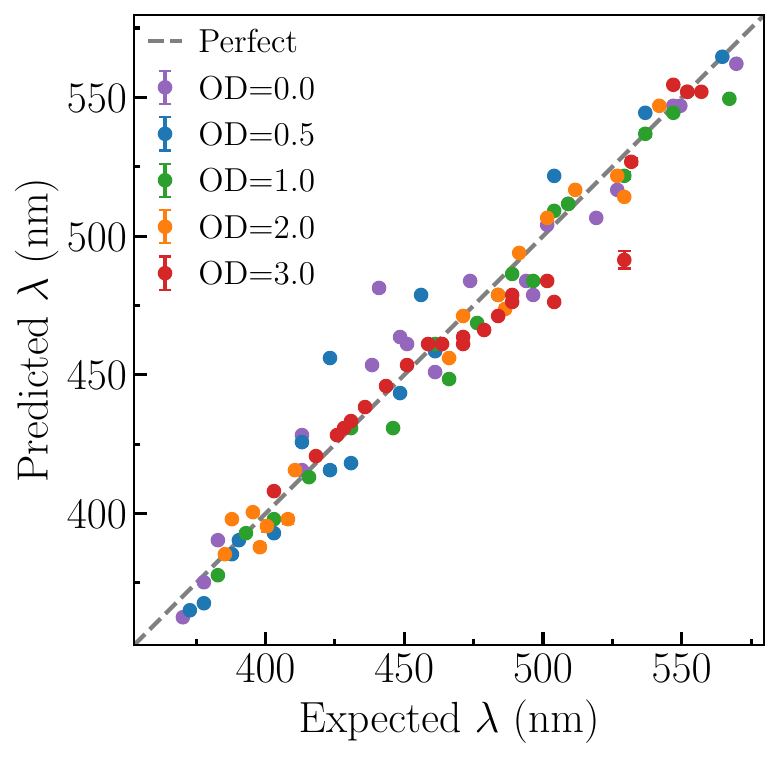}\label{fig:parity:wavelength}}%
    \hspace{0.5em}%
    \subfloat[\vspace{-0.75em}]{\includegraphics[scale=0.3775]{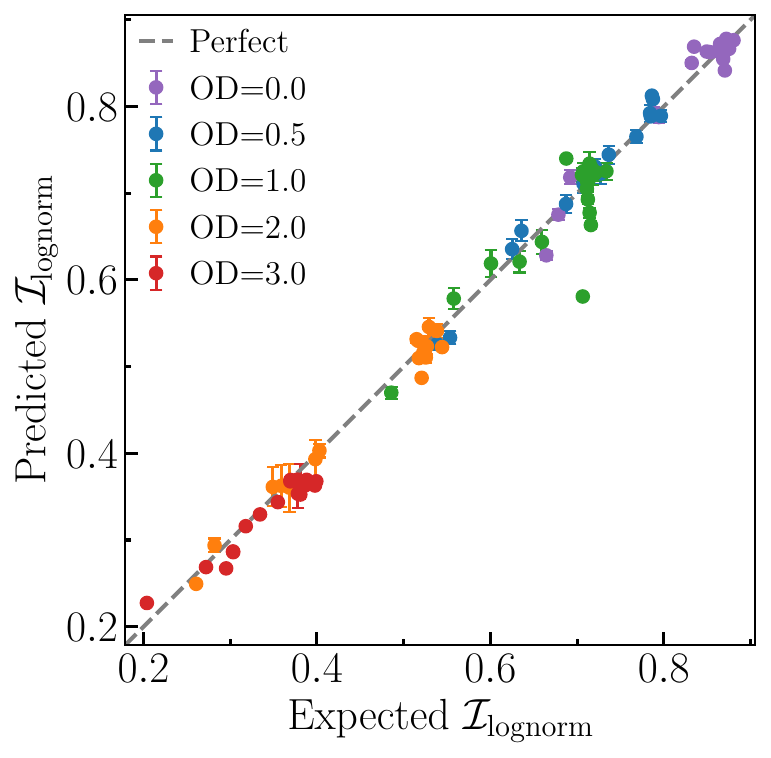}\label{fig:parity:log_irradiance}}%
    \hspace{0.5em}%
    \subfloat[\vspace{-0.75em}]{
    \includegraphics[scale=0.3775]{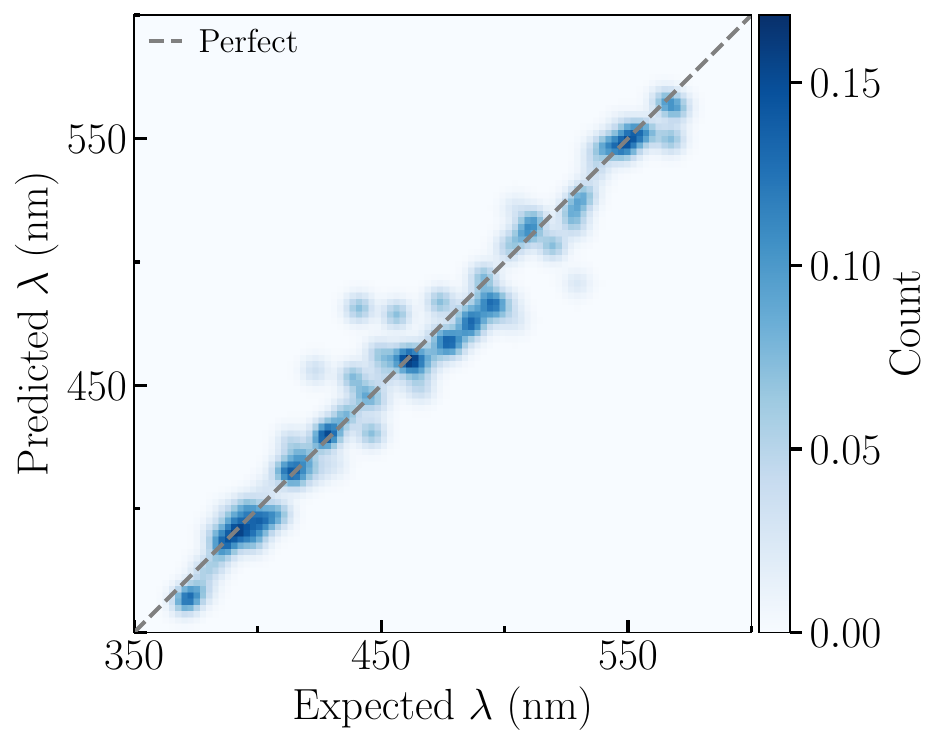}\label{fig:parity:wavelength_confusion}}
    \caption{%
        \textbf{Predictive performance of the spectral reconstruction model.}
        Reconstructed versus experimental \protect\subref{fig:parity:wavelength} wavelength and \protect\subref{fig:parity:log_irradiance} log-normalised irradiance for the held-out test set.
        The dashed line denotes ideal prediction; colours indicate optical-density filter (OD) level and vertical whiskers show the spread of the latent-space ensemble.
        \protect\subref{fig:parity:wavelength_confusion} Prediction-density map of reconstructed versus experimental wavelength, highlighting the wavelength dependence of the reconstruction accuracy.%
    }
    \label{fig:parity}
\end{figure}

Having established the structure of the inverse-learning framework, we next test whether the nonlinear electrical encoding can be inverted for optical conditions not encountered during training.
\figrefs{fig:parity:wavelength}{fig:parity:log_irradiance} compares the reconstructed wavelength and log-normalised irradiance with their corresponding experimental values for the held-out test set of 95 voltage sweeps, with test wavelength-irradiance pairs explicitly excluded from model training.
The wavelength predictions closely follow the ideal parity relation across the investigated spectral range, yielding $R^2=0.958$, a mean absolute error of $8.1$~\si{\nano\metre} and an RMSE of $11.2$~\si{\nano\metre}.
Irradiance is reconstructed with still higher fidelity, reaching $R^2=0.987$ in log-normalised space and approximately $R^2=0.96$ when expressed on the original linear scale (see Supporting Information
Section~S6.1
for the corresponding parity plot of irradiance in the linear scale).
The predictions shown in \figref{fig:parity} are obtained using the 100-member latent-space ensemble described above; across these realisations, $R^2=0.9584\pm0.0103$ for wavelength and $R^2=0.9870\pm0.0048$ for log-normalised irradiance.
Using the nominal inference for deterministic inference (instead of the ensemble sampling) yields marginally higher $R^2$ values ($0.9588$, $0.9871$, and $0.9591$ for wavelength, log-normalised irradiance, and linear irradiance, respectively; see Supporting Information
Section~S6.2%
), though the ensemble approach is preferred as it additionally provides a measure of prediction sensitivity.
These results demonstrate that the learned inverse mapping generalises beyond the optical states used for optimisation, allowing both wavelength and irradiance to be recovered from the nonlinear response of a single photodetector.

\begin{figure}
    \centering%
    \subfloat[\vspace{-0.75em}]{\includegraphics[height=0.40\linewidth]{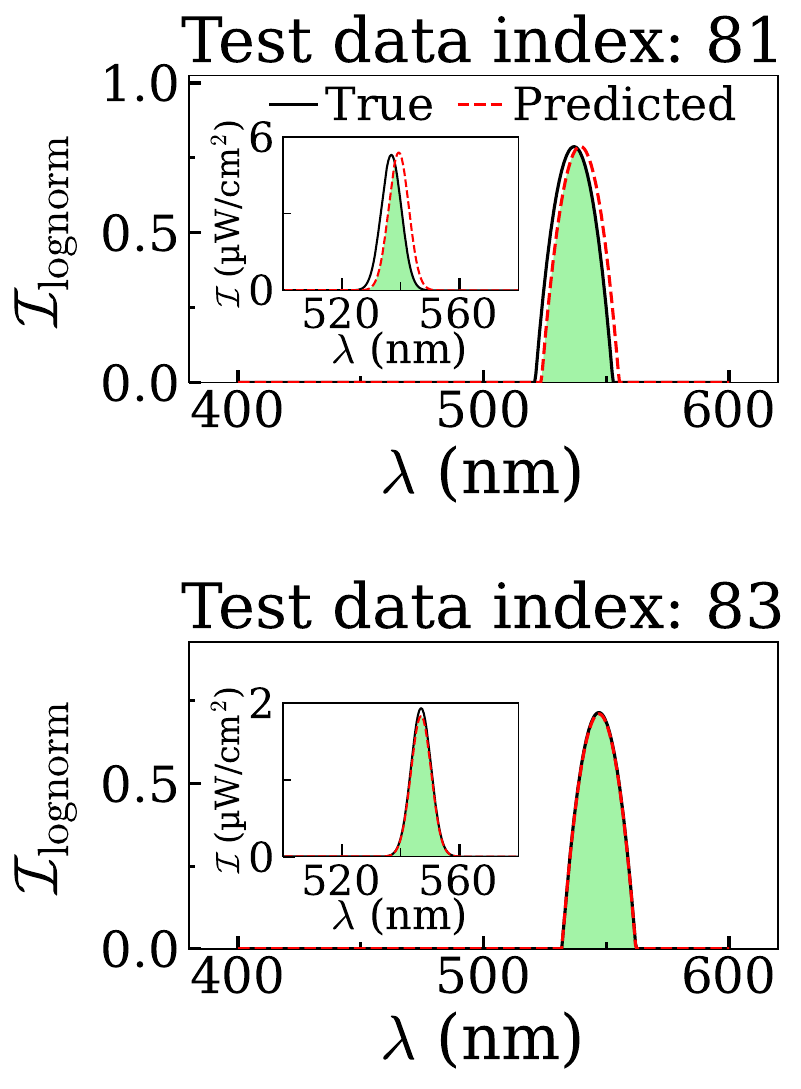}\label{fig:pred:a}}%
    \hspace{0.5em}%
    \subfloat[\vspace{-0.75em}]{\includegraphics[height=0.40\linewidth]{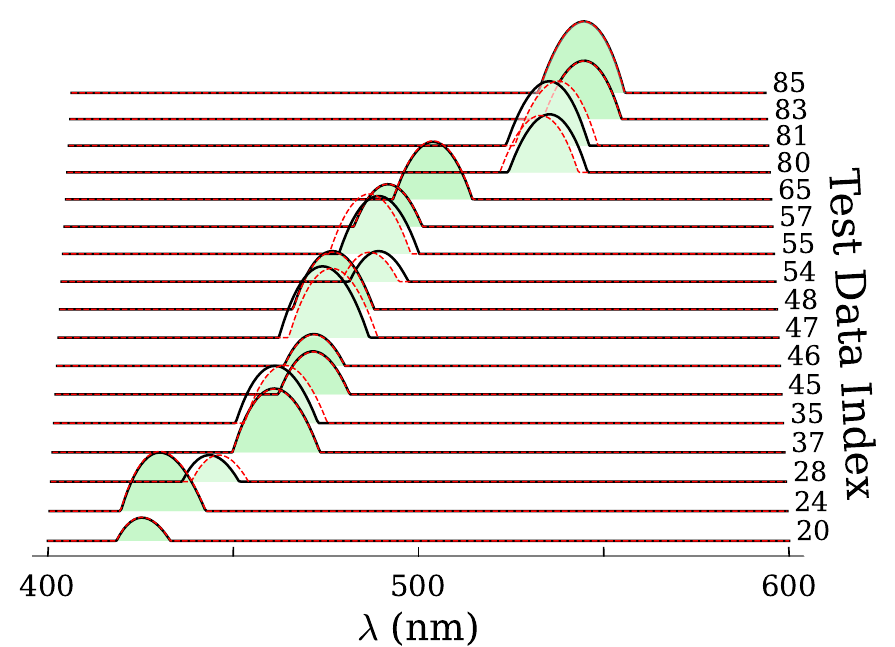}\label{fig:pred:b}}%
    \caption{%
        \textbf{Representative spectral reconstructions from the held-out test set.}
        \protect\subref{fig:pred:a}~Experimental input spectra (True, black line) and corresponding model reconstructions (Predicted, red stipulated line) are shown for representative combinations of wavelength and irradiance, illustrating recovery of both spectral position and amplitude across the investigated optical range, the insets show the true and predicted wavelength (\si{\nano\metre}) and irradiance in \si{\micro\watt\per\centi\metre\squared}.
        \protect\subref{fig:pred:b}~A larger subset of the test data showing input spectra (True, black line) and corresponding model reconstructions (Predicted, red stipulated line). Due to the large range of irradiances in the dataset, plotted irradiance data is shown in log$_{10}$ scaling (apart from insets in \protect\subref{fig:pred:a}). 
    }
    \label{fig:reconstructed_spectra}
\end{figure}

The reconstruction error is not uniformly distributed across wavelength, providing insight into the limits imposed by the physical encoding itself.
The prediction-density map in \figref{fig:parity:wavelength_confusion} is tightly concentrated around the ideal relation below approximately $425$~\si{\nano\metre} and above approximately $530$~\si{\nano\metre}, whereas a broader distribution develops across the intermediate $425$-$530$~\si{\nano\metre} region.
This increased ambiguity coincides with the non-monotonic spectral responsivity of the F-PEAI detector in \figref{fig:device:photoresponsivity}, where distinct wavelengths can produce comparable responsivities at a single operating bias.
Such degeneracy reduces the separability of neighbouring optical states and consequently makes the inverse problem more demanding.
Importantly, accurate reconstruction remains possible across this region because the model does not rely on a single responsivity value, but decodes the full nonlinear and history-dependent current-voltage signature.
To illustrate the reconstruction directly, \figref{fig:reconstructed_spectra} compares representative predicted spectra with their corresponding optical inputs across the test set.
The reconstructed peaks closely reproduce both the position and amplitude of the incident excitation over widely separated regions of the wavelength-irradiance space, providing a direct spectral representation of the performance quantified statistically in \figref{fig:parity}.
The residual wavelength dependence of the reconstruction error therefore suggests that reconstruction performance is constrained not only by the numerical model, but also by the information content and separability of the physical encoding.

The recovery of irradiance is particularly revealing.
Before entering the encoder, the current traces are normalised by their maximum absolute magnitude, removing the raw absolute photocurrent scale that would otherwise provide a direct measure of illumination strength.
Nevertheless, the model reconstructs log-normalised irradiance with $R^2=0.987$.
Irradiance information must therefore persist in the internal shape of the normalised electrical response rather than being encoded solely in its absolute amplitude.
Moreover, the model's internal decomposition process inherently suppresses noise, meaning that irradiance cannot simply be decoded from the signal-to-noise ratio of the input.

The interpretation that irradiance is encoded in the response shape is further supported by an additional observation: removing the dark current input feature leaves irradiance performance largely intact (see Supporting Information
Section~S7.5%
).
Instead, the paired corrected-photocurrent and dark-current channels retain relative information that evolves across the voltage sweep and can be decoded by the network; the broadly distributed correlation with irradiance observed across the sweep provides further support for this interpretation (see Supporting Information
\figurename~S6%
).
So, whilst the paired corrected-photocurrent and dark-current channels retain relative information that can be decoded by the network, the majority of this information is encoded in the shape of the photocurrent response.

The simultaneous reconstruction of wavelength and irradiance therefore exploits different aspects of the same nonlinear electrical trajectory: wavelength is encoded predominantly through changes in response morphology and bias dependence, whereas irradiance remains accessible through more distributed relationships within the normalised detector response.

The model also generalises across all investigated optical-density filter levels (see Supporting Information
Section~S6.3%
), with wavelength reconstruction yielding $R^2=0.922$-$0.981$ despite the relatively small number of test samples available for each attenuation level; only a modest reduction in performance is observed at the highest attenuation, optical-density filter (OD) $3.0$, where the signal-to-noise ratio is lowest.

We further tested whether this predictive performance is robust to dataset size, model initialisation, unseen wavelengths, latent-space sampling, input features, and model architecture (see Supporting Information
Section~S7
for a detailed discussion on each of these points).
Increasing the fraction of experimental data used for training from $10\%$ to $80\%$ produces an overall improvement in reconstruction accuracy, without a clear performance plateau, indicating that the present model remains data-limited and could benefit from larger experimental datasets.
Across 20 random seeds (used to vary both model initialisation and data partitioning), performance remains highly reproducible, with $R^2=0.9463\pm0.0115$ for wavelength and $R^2=0.9840\pm0.0032$ for log-normalised irradiance; see Supporting Information
\figurename~S9
and
Table~S2%
.
Even when entire wavelengths are withheld from training, the model maintains strong reconstruction performance ($R^2=0.9191$).
Replacing the learned latent noise scale, typically 
$\sigma\approx0.04$, with a fixed value of $0.5$ increases the spread of the ensemble predictions by approximately an order of magnitude while changing $R^2$ by less than $0.01$.
The model performs best with the input features of photocurrent, dark current, and voltage range; omitting voltage range slightly reduces model performance, whilst removing dark current significantly decreases model performance across all metrics.
Finally, for model training and architecture, omitting pretraining or averaging over the forward and reverse voltage sweeps significantly impact model performance.
Together, these tests show that the reconstruction is stable against stochastic initialisation and unseen wavelengths, the chosen architecture and feature inputs are necessary to properly decode spectral data from photoresponse, while the absence of data saturation suggests that further gains should be achievable through expansion of the training dataset.

\subsection{Voltage-resolved encoding of spectral information}

\begin{figure}[t]
    \centering
    \subfloat[\vspace{-0.75em}]{\includegraphics[scale=0.425]{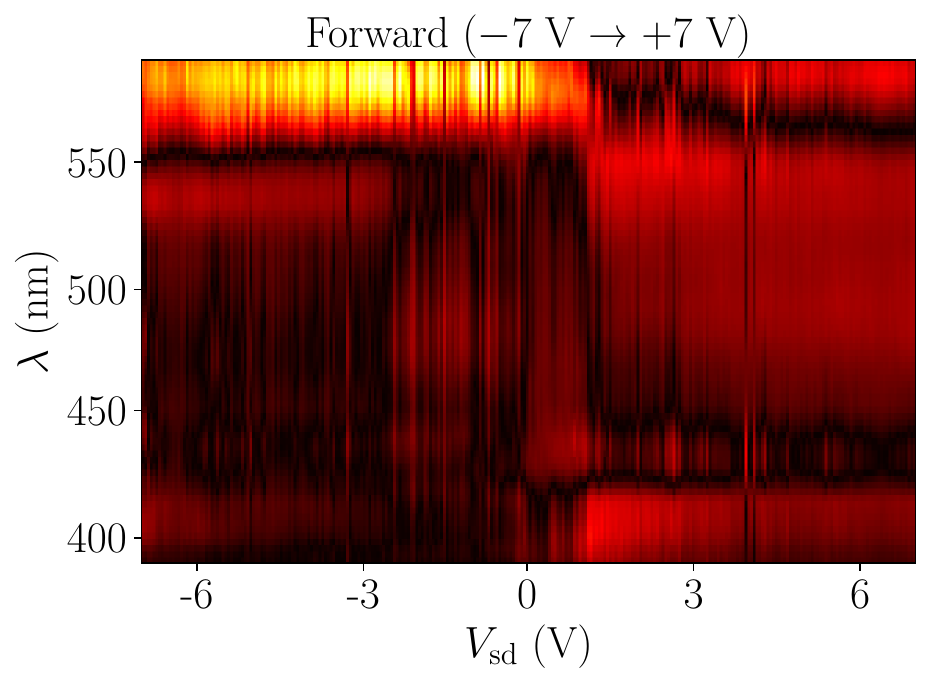}\label{fig:correlation:forward}}%
    \hspace{0.5em}%
    \subfloat[\vspace{-0.75em}]{\includegraphics[scale=0.425]{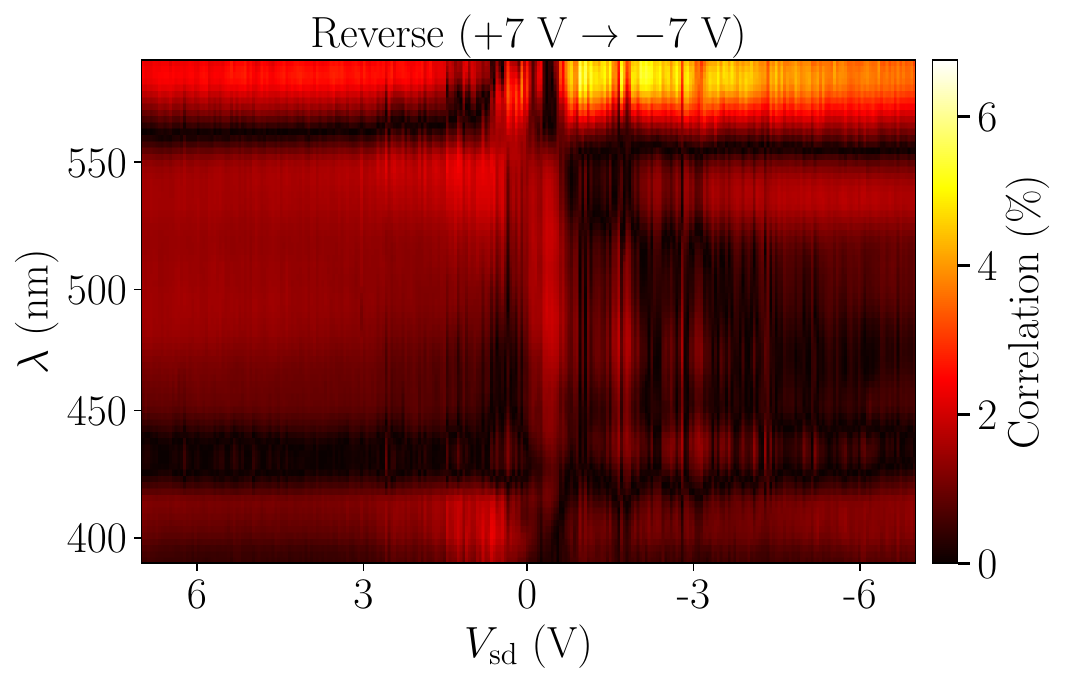}\label{fig:correlation:reverse}}%
    \vspace{-1em}\hspace{0.5em}%
    \subfloat[\vspace{-0.75em}]{\includegraphics[scale=0.425]{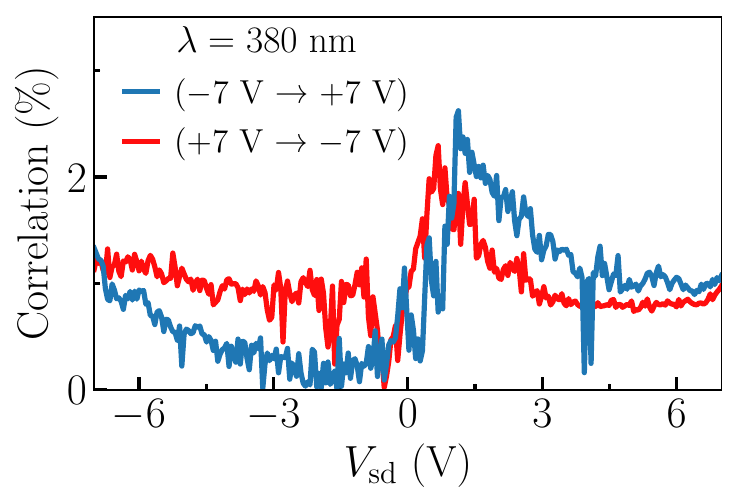}\label{fig:correlation:380}}%
    \hspace{0.5em}%
    \subfloat[\vspace{-0.75em}]{\includegraphics[scale=0.425]{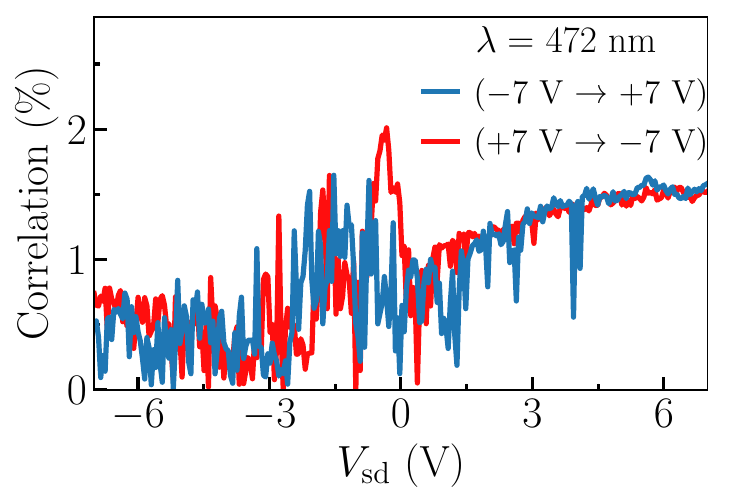}\label{fig:correlation:472}}%
    \hspace{0.5em}%
    \subfloat[\vspace{-0.75em}]{\includegraphics[scale=0.425]{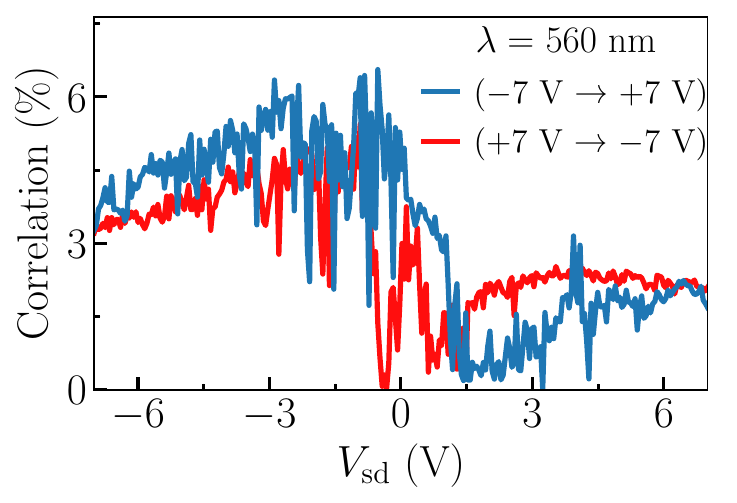}\label{fig:correlation:560}}%
    \caption{%
        \textbf{Voltage-resolved encoding of wavelength information in the nonlinear photocurrent response.}
        Row-normalised absolute Pearson correlation between photocurrent and experimental spectral intensity as a function of source-drain voltage and wavelength for the \protect\subref{fig:correlation:forward} forward and \protect\subref{fig:correlation:reverse} reverse bias sweeps.
        Representative voltage-dependent correlation profiles at \protect\subref{fig:correlation:380} 380, \protect\subref{fig:correlation:472} 472, and \protect\subref{fig:correlation:560} 560~\si{\nano\metre}, comparing the forward and reverse sweep directions.
        The complete correlation analysis and definitions are provided in Supporting Information Sections S5 and S6.4.}
    \label{fig:correlation}
\end{figure}

The successful spectral reconstruction raises a more fundamental question: how is optical information distributed across the nonlinear electrical response?
To address this, we calculate the absolute Pearson correlation between the photocurrent at each source-drain voltage and the experimental spectral intensity at each wavelength, resolving the forward and reverse sweep directions independently (\figrefs{fig:correlation:forward}{fig:correlation:reverse}; Supporting Information Sections S5 and S6.4)).
For visualisation, the correlation matrices are normalised across wavelength at each voltage, revealing a strongly structured wavelength- and bias-dependent distribution rather than a uniform association across the current-voltage trajectory.
Representative slices at 380, 472 and 560~\si{\nano\metre} (\figrefspan{fig:correlation:380}{fig:correlation:560}) expose a pronounced evolution of this structure: below approximately 425~\si{\nano\metre}, the relative correlation weight is distributed comparably across positive and negative bias, with a modest enhancement at positive bias; between approximately 425 and 545~\si{\nano\metre}, it becomes preferentially concentrated at positive bias; above approximately 545~\si{\nano\metre}, this balance reverses and the negative-bias response becomes dominant.
Forward and reverse sweeps also exhibit distinct correlation landscapes, consistent with the history-dependent information retained by their independent encoding.
This analysis should not be interpreted as a causal attribution of individual voltage points to the neural-network prediction; rather, it identifies where the measured electrical response co-varies most strongly with different regions of the experimental spectrum (the equivalent plots for correlation of electrical response to reconstructed spectrum is presented in the Supporting Information
Section~S6.4%
, but shows very little difference with the experimental correlation, as expected due to the high wavelength $R^2$).
Irradiance exhibits qualitatively different behaviour, with its correlation distributed more broadly across both bias polarities and sweep directions (Supporting Information
Section~S6.4%
).
Wavelength and irradiance therefore exhibit distinct correlation structures within the same nonlinear detector response, supporting the conclusion that different regions of the device operating landscape carry complementary optical information.

\section{Conclusions}

In conclusion, we demonstrate that the nonlinear optoelectronic dynamics of a single 2D perovskite photodetector can serve as a computational resource for spectroscopy, enabling wavelength and irradiance to be reconstructed from electrically measured current-voltage trajectories without dispersive optics or detector arrays.
Rather than suppressing the trapping, interfacial transport, bias asymmetry and history dependence conventionally regarded as limitations of photodetectors, our approach exploits the multidimensional response generated by these processes as an information-rich physical encoding of the incident optical field.
The structure-aware variational architecture complements this physical encoder by preserving the functional and hysteretic character of the response and decoding it into continuous spectral parameters.
Crucially, the voltage-resolved analysis shows that wavelength and irradiance are encoded differently within the same response: wavelength information is distributed across distinct bias regions (short wavelengths $<425$~\si{\nano\metre} across both polarities, mid-wavelengths $425$-$545$~\si{\nano\metre} at positive bias, and long wavelengths $>545$~\si{\nano\metre} at negative bias), whereas irradiance exhibits a more uniform correlation across bias.
The model reconstructs wavelength with $R^2=0.96$ (MAE $=8.1$~\si{\nano\metre}, RMSE $=11.2$~\si{\nano\metre}) and log-scaled irradiance with $R^2=0.99$; this performance remains reproducible across 20 random seeds ($R^2=0.9463\pm0.0115$ for wavelength, $0.9840\pm0.0032$ for irradiance), whilst latent-space sampling additionally provides per-prediction uncertainty estimates.
Ablation studies confirm that retaining dark current as a model input is critical for wavelength reconstruction.
The compact inference architecture ($14,799$ parameters) is compatible with deployment on modest computational resources without specialised hardware, aligning with the broader vision of accessible, hardware-algorithm co-designed sensing systems proposed for next-generation computational photodetectors~\cite{Wang20252DComputationalPhotodetectors}.
This work establishes a route towards hardware-algorithm co-design in which materials, interfaces and device operating conditions are engineered together with the inference architecture to maximise the separability and information content of the measured response, rather than optimising detector responsivity alone.
More broadly, such an approach could transform nonlinearities and internal dynamics from imperfections to be eliminated into functional degrees of freedom for computation and sensing, opening a path towards compact spectroscopic systems in which material physics performs the encoding and machine learning provides the inverse decoding.

\section*{Methods}

\subsection*{Fabrication of contacts}
Quartz substrates (500~\si{\micro\metre} thick) were spin-coated at 2000~\si{\rpm} with a $\sim$400~\si{\nano\metre}-thick layer of 450K A6 PMMA and baked at 180~\si{\celsius} for 3~\si{\minute}.
To mitigate charging during electron-beam lithography, a 20~\si{\nano\metre} Al conductive layer was deposited by electron-beam evaporation. Electrodes were patterned by 100~\si{\kilo\electronvolt} electron-beam lithography at a dose of 1000~\si{\micro\coulomb\per\centi\metre\squared}. Before development, the Al layer was removed by immersion in MF319 for 60~\si{\second}, and the PMMA was developed for 90~\si{\second} in IPA:MIBK (3:1 by volume). Contacts were metallised by electron-beam evaporation of Ti/Au (5/30~\si{\nano\metre}), followed by lift-off in acetone.

\subsection*{\texorpdfstring{Synthesis of (F-PEA)$_{2}$PbI$_{4}$}{Synthesis of (F-PEA)2PbI4}}
Single crystals of F-PEAI were grown at room temperature by antisolvent vapour-assisted crystallisation following a previously optimised protocol; their single-crystalline phase has been established by synchrotron X-ray diffraction~\cite{Mastria2024RealTimeHighly}. Briefly, 267~\si{\milli\gram} of 4-fluorophenethylammonium iodide and 230.5~\si{\milli\gram} of PbI$_2$ were dissolved in 1~\si{\milli\litre} of $\gamma$-valerolactone inside a nitrogen glovebox and stirred at 70~\si{\celsius} for 30~\si{\minute}. A 2~\si{\micro\litre} droplet of the precursor solution was confined between two cleaned glass substrates and exposed to dichloromethane vapour in a sealed Teflon chamber for 12~\si{\hour}, yielding high-quality single crystals.

\subsection*{\texorpdfstring{Exfoliation and transfer of (F-PEA)$_2$PbI$_4$}{Exfoliation and transfer of (F-PEA)2PbI4}}
Thin F-PEAI flakes were obtained from bulk single crystals by mechanical exfoliation using thermal-release tape (Graphene Supermarket, SKU: GTT-5P), following previously established procedures~\cite{Mastria2024RealTimeHighly,Riisnaes20242DHybridPerovskite}. Flakes were inspected by optical microscopy, and those exhibiting uniform optical contrast indicating consistent thickness and sufficiently large lateral dimensions were selected for device fabrication. The transparency of the thermal-release tape enabled optical alignment of the selected flakes with prefabricated Au electrode patterns. Following alignment, the tape was brought into contact with the substrate and the assembly heated to 95~\si{\celsius} for 10~\si{\second}, releasing the adhesive and transferring the F-PEAI flake onto the electrodes; see Supporting Information Section S1.

\subsection*{Spectroscopic and opto-electronic characterisation}
Optical absorption and photoluminescence (PL) measurements were performed using a custom-built multimodal optoelectronic characterisation platform developed for atomically thin materials and hybrid optoelectronic devices~\cite{DeSanctis2017IntegratedMultiPurpose}. Excitation was provided by digitally controlled continuous-wave solid-state lasers (Coherent OBIS 375LX, 473LS, 514LX and 561LS; Omicron LuxX 685), with output powers of 30-50~\si{\milli\watt}. Laser intensity was continuously controlled by analogue power modulation, while custom-designed filter holders enabled rapid insertion of optical-density, band-pass and notch filters, as well as polarisation optics, without altering the optical alignment.

Spectral analysis was performed using a Princeton Instruments Acton SP2500 imaging spectrometer equipped with interchangeable diffraction gratings (1200~\si{\gram\per\milli\metre} with 500 and 750~\si{\nano\metre} blaze wavelengths; 1800~\si{\gram\per\milli\metre} with a 500~\si{\nano\metre} blaze wavelength) and a thermoelectrically cooled Princeton Instruments PIXIS400-eXcelon back-illuminated CCD detector. The modular optical configuration enabled rapid switching between Raman spectroscopy, photoluminescence, transmission/reflection spectroscopy and photocurrent mapping through appropriate selection of the optical filters.

Bias-dependent photocurrent measurements were acquired using a Newport TLS300X xenon-lamp monochromator system. The incident optical power was varied using calibrated optical-density filters mounted on motorised, computer-controlled filter wheels, and the irradiance at the sample plane was measured for each optical configuration using a calibrated Thorlabs S130CV silicon photodiode. Source-drain bias was applied using a Xitron 2000 precision current-voltage source. The photocurrent was amplified using an Ithaco 1211 low-noise current preamplifier and recorded with Agilent 34401A digital multimeters.

\subsection*{Data}

\textbf{Data preprocessing.}
Prior to machine-learning training, an automated quality-control procedure was applied to remove voltage sweeps exhibiting insufficient or noise-dominated bias-dependent modulation.
For this procedure, the source-drain voltage and photocurrent response of each sweep were temporarily normalised by their respective maximum absolute values, and the forward and reverse branches were independently fitted with first-order polynomials.
Sweeps for which either branch exhibited a normalised slope below 0.5, corresponding to less than a 50\% variation in normalised responsivity across the applied bias range, were discarded.
This procedure removed 29 low-information sweeps from the original dataset of 505 measurements, leaving 476 sweeps for model development and evaluation.

Irradiance, $\mathcal{I}$, was logarithmically normalised according to

\begin{equation}
\mathcal{I}_{\textrm{lognorm}} =
\frac{\log{\mathcal{I}}-\log{\mathcal{I}_{\textrm{min}}}}
{\log{\mathcal{I}_{\textrm{max}}}-\log{\mathcal{I}_{\textrm{min}}}},
\end{equation}

\noindent
where $\mathcal{I}_{\textrm{min}}=2.\dot{2}\times10^{-6}$ and $\mathcal{I}_{\textrm{max}}=2.\dot{2}$~\si{\watt\per\centi\metre\squared}, corresponding to optical powers of $10^{-10}$ and $10^{-4}$~\si{\watt}, respectively, for a device active area of $4.5\times10^{-5}$~\si{\centi\metre\squared}.
This range is set to encompass the entire irradiance range without clipping it to the exact extent of the available dataset.
This logarithmic transformation accommodates the large dynamic range in irradiance while mapping the prescribed interval onto a dimensionless scale between 0 and 1.

\textbf{Data validity.}
After removal of high-noise measurements, the retained data remain distributed across the investigated wavelengths and optical-density-filter conditions, providing coverage of the detector's nonlinear response across the explored dynamic range.
Model inputs comprise the source-drain voltage, normalised corrected photocurrent and normalised dark current.
To limit overfitting within the comparatively small experimental dataset, the architecture incorporates polynomial decomposition, variational augmentation and latent-space regularisation, with test wavelengths-irradiance pairs explicitly excluded from model training.

Further details of the data preprocessing and normalisation procedures are provided in Supporting Information
Section~S2%
.

\subsection*{Machine Learning Model Architecture}
\label{sec:methods:model}

We employ a variational autoencoder(VAE)-inspired variational encoder-decoder comprising an analytical Legendre-polynomial representation, a probabilistic neural-network encoder and a Gaussian-peak decoder that maps the measured electrical response onto the incident spectrum.
The model takes as inputs the source-drain voltage, corrected photocurrent and dark-current traces and predicts the amplitude, centre wavelength and width of $K$ Gaussian spectral components; $K=1$ is used for the monochromatic spectra considered here.

\textbf{Encoder: Legendre polynomial decomposition.}
Each measurement comprises an increasing-bias sweep followed by a decreasing-bias sweep.
To preserve directional and hysteretic information, the forward and reverse branches are treated independently.
For each sweep direction $s\in\{\rightarrow,\leftarrow\}$, the source-drain voltage is mapped onto the dimensionless interval $V_{\mathrm{norm}}\in[-1,1]$.
Two input-current channels $f$ are considered independently: the corrected photocurrent and the dark current.
Writing $t$ for the sampled points along each sweep, each current trace $I_{f,s}(t)$ is projected onto a Legendre-polynomial basis truncated at maximum degree $D=8$:

\begin{equation}
I_{f,s}(t)\simeq\sum_{d=0}^{D}c_{d,f,s}P_d\left(V_{\mathrm{norm}}(t)\right),
\end{equation}

\noindent
where $P_d$ is the Legendre polynomial of degree $d$. The coefficients are obtained by orthogonal projection,

\begin{equation}
c_{d,f,s}=
\frac{\left\langle I_{f,s},P_d\right\rangle_t}
{\left\langle P_d,P_d\right\rangle_t}.
\end{equation}

\noindent
The resulting coefficients are concatenated with the minimum and maximum sweep voltages, $V_{\mathrm{min}}=-7$~\si{\volt} and $V_{\mathrm{max}}=+7$~\si{\volt}, thereby retaining the physical voltage scale after normalisation.
The feature vector is batch-normalised and passed through a fully connected layer with 128 hidden units, followed by a further batch normalisation, a leaky-ReLU activation, and dropout (dropout is disabled during inference).
Two parallel linear heads (single fully connected layer with 32 hidden units and learnable bias parameters) predict the latent mean $\boldsymbol{\mu}\in\mathbb{R}^{32}$ and log-variance $\log\boldsymbol{\sigma}^{2}\in\mathbb{R}^{32}$, defining the approximate posterior

\begin{equation}
q_{\phi}(\mathbf{z}|\mathbf{x})=
\mathcal{N}\left(\boldsymbol{\mu},
\operatorname{diag}(\boldsymbol{\sigma}^{2})\right).
\end{equation}

\noindent
During training, latent states are sampled using the reparameterisation trick~\cite{Kingma2013AutoEncodingVariational,Rezende2014StochasticBackpropagationApproximate},
$\mathbf{z}=\boldsymbol{\mu}+\boldsymbol{\sigma}\odot\boldsymbol{\epsilon}$, with $\boldsymbol{\epsilon}\sim\mathcal{N}(\mathbf{0},\mathbf{I})$, enabling gradient propagation through the stochastic latent layer.
Nominal inference uses $\mathbf{z}=\boldsymbol{\mu}$, while repeated sampling from the learned posterior generates the ensemble used to assess the sensitivity of the spectral reconstruction.
Each fully connected layer in the encoder includes a learnable bias parameters (additive offsets, not to be confused with the input voltage bias).
The encoder contains 13,580 trainable parameters.

\textbf{Decoder: Gaussian peak prediction.}
The decoder maps the latent vector $\mathbf{z}$ onto the physical parameters of $K$ Gaussian spectral components: amplitude $A_k$, centre wavelength $\mu_k$ and width $\sigma_k$.
The latent representation is first passed through a shared fully connected layer with 16 units, batch normalisation and a leaky-ReLU activation, and dropout (disabled during inference), before branching into two specialised heads.
The position head predicts $\mu_k$ through a sigmoid-activated output scaled to the operating wavelength range, while the amplitude-width head predicts the logarithmic amplitude and $\sigma_k$.
Both heads follow the same internal structure: a fully connected layer with 16 units, batch normalisation, a leaky-ReLU activation, dropout (disabled during inference), and a final layer that projects onto the required output dimensions ($K$ for position head and $K\times 2$ for the amplitude-width head).
The log-amplitude is constrained through a sigmoid-scaled transformation to the prescribed range $[10^{-10},10^{-4}]$. The reconstructed spectrum is

\begin{equation}
S(\lambda)=\sum_{k=1}^{K}
\frac{A_k}{\sqrt{2\pi\sigma_k^2}}
\exp\left[-\frac{(\lambda-\mu_k)^2}{2\sigma_k^2}\right],
\label{eq:reconstruction}
\end{equation}

\noindent
where $\lambda$ is the wavelength and $K$ is the maximum number of spectral components.
The present study uses $K=1$.
Because the experimental linewidth is fixed across the dataset, $\sigma_k$ is retained as a physically meaningful model parameter but is not independently benchmarked.
Each fully connected layer in the decoder includes a learnable bias parameters (additive offsets, not to be confused with the input voltage bias).
The decoder contains 1,219 trainable parameters.
The complete model contains 14,799 trainable parameters.

\textbf{Training and pretraining.}
Training proceeds in two stages. The encoder is first pretrained using the self-supervised SimSiam objective~\cite{Chen2021ExploringSimpleSiamese}.
Two augmented views of each unlabelled electrical response are generated using Fourier-domain phase perturbations and low-frequency masking, and their latent representations are driven towards consistency by maximising cosine similarity under a stop-gradient operation.
A simultaneous regularisation term penalises deviations of the latent distribution from a unit Gaussian (Kullback-Leibler, KL, divergence)~\cite{Kingma2013AutoEncodingVariational,Rezende2014StochasticBackpropagationApproximate}.
The pretraining objective combines the similarity and latent-regularisation terms with weights $\lambda_{\mathrm{simsiam}}=0.75$ and $\lambda_{\mathrm{KL}}=1.0$, respectively, and is optimised for 50 epochs. Further details are provided in Supporting Information
Section~S3%
.

The complete encoder-decoder is subsequently trained under supervision against the parameters of the target Gaussian spectrum.
A differentiable soft-assignment procedure based on a temperature-annealed softmax over pairwise peak-position distances establishes correspondences between predicted and target spectral components.
The total loss combines contributions from peak position, amplitude, width and minimum-amplitude regularisation, with weights $\lambda_{\mathrm{pos}}=1.0$, $\lambda_{\mathrm{amp}}=1.2$, $\lambda_{\mathrm{sigma}}=0.2$ and $\lambda_{\mathrm{min-amp}}=0.3$, respectively. Width errors are evaluated only for sufficiently confident assignments, while the amplitude terms suppress superfluous spectral components. Supervised optimisation converges within 600 epochs. The present study uses a single predicted spectral component ($K=1$), while the multi-peak formulation is detailed in Supporting Information
Section~S4%
.

Both training stages use AdamW with a learning rate of $10^{-2}$ and a batch size of 128.
Training uses 80\% of the accepted experimental measurements, with the remaining 20\% of the data reserved for testing as described above.
Hyperparameter optimisation and experiment tracking were performed using Weights \& Biases \cite{wandb}.
Independent implementations in PyTorch and ATHENA~\cite{Taylor2024ATHENAFortranPackage} produced quantitatively equivalent results, subject to differences arising from random-number generation.

\textbf{Inference.}
Predictions reported in the main text are obtained from an ensemble of 100 latent-space samples for each input measurement.
Each sample is drawn from the learned approximate posterior and propagated through the Gaussian decoder.
The ensemble mean defines the reported spectral prediction, while the spread across the ensemble provides the model-internal sensitivity measure discussed in the main text.

\section*{Data availability}
All data supporting the conclusions of this study are provided in the main text and/or the Supporting Information. Additional data and code are available from the corresponding authors upon reasonable request.

\section*{CRediT author statement}

N. T. T. contributed to investigation, methodology, formal analysis, software, visualisation, project administration, data curation and funding acquisition.
K. J. R. fabricated the 2D F-PEAI photodetector, acquired the initial photocurrent and spectral-responsivity measurements, performed data analysis and visualisation, data curation and developed an independent version of the machine-learning algorithm.
R. M. and F. S. D. synthesised the 2D F-PEAI single crystals.
H.O. L. performed the optical absorption and photoluminescence measurements.
S. R. and M. F. C. conceived the project.
S. R. acquired the experimental training dataset, developed an initial independent machine-learning neural network algorithm, and supervised the progress of the project.
N. T. T. and S. R. wrote the original manuscript, with all authors contributing to the review and editing of sections relevant to their expertise.

\section*{Acknowledgement}
For this work, N.T.T was supported by the Government Office for Science and the Royal Academy of Engineering under the UK Intelligence Community Postdoctoral Research Fellowships scheme (Grant No. ICRF2425-8-148).
K.J.R, M.F.C. and S.R. acknowledge financial support from the Leverhulme Trust (grants "Graded excitonics" and "Giant Permittivity"),  EPSRC (Grant no. EP/K010050/1, EP/M001024/1, EP/M002438/1, EP/Z534250/1, EP/Y021339/1, EPSRC Core Equipment 2024-UKRI323), and EU H2020-MSCA-RISE projects TERASSE (Project No. 823878), CHARTIST (Project No. 101007896) and HERMES (Project No. Project 101236439).
R.M. acknowledge financial support from the project AI-PHOQUS "Artificial Intelligence and Advanced Networks Embedded in Photonics and Quantum Sciences and Technology", CUP B83C26000470007 within the "NP Research, innovation and competitiveness for green and digital transition 2021-2027" (PN RIC 2021-2027) co-financed by the European Union, the Italian Ministry of Enterprises and Made in Italy (MIMIT) and the Italian Ministry of University and Research (MUR).

\bibliographystyle{unsrt}
\bibliography{references}

\end{document}


\title{Supporting Information: Compact Variational Neural Networks for Spectral Inference from a Single Nonlinear 2D Perovskite Photodetector}

\author{Karl Jonas Riisnaes}
\altaffiliation{These authors contributed equally to this work.}
\affiliation{Centre for Graphene Science, Department of Physics and Astronomy, University of Exeter, Exeter EX4 4QL, United Kingdom}

\author{Ned Thaddeus Taylor}
\altaffiliation{These authors contributed equally to this work.}
\affiliation{Centre for Graphene Science, Department of Physics and Astronomy, University of Exeter, Exeter EX4 4QL, United Kingdom}

\author{Hoi Tung Lam}
\affiliation{Centre for Graphene Science, Department of Physics and Astronomy, University of Exeter, Exeter EX4 4QL, United Kingdom}

\author{Rosanna Mastria}
\affiliation{CNR NANOTEC, Institute of Nanotechnology, Via Monteroni, 73100 Lecce, Italy}

\author{Francesco Saverio Difeo}
\affiliation{CNR NANOTEC, Institute of Nanotechnology, Via Monteroni, 73100 Lecce, Italy}
\affiliation{Department of Mathematics and Physics “Ennio de Giorgi”, University of Salento, via Monteroni, 73100, Lecce, Italy}

\author{Monica Felicia Craciun}
\affiliation{Centre for Graphene Science, Department of Engineering, University of Exeter, Exeter EX4 4QL, United Kingdom}

\author{Saverio Russo}
\email{s.russo@exeter.ac.uk}
\affiliation{Centre for Graphene Science, Department of Physics and Astronomy, University of Exeter, Exeter EX4 4QL, United Kingdom}

\maketitle

\tableofcontents
\listoffigures
\listoftables

\clearpage{}
\section{Device}
\label{sec:device}

\begin{figure}[H]
    \centering
    \includegraphics[width=0.9\linewidth]{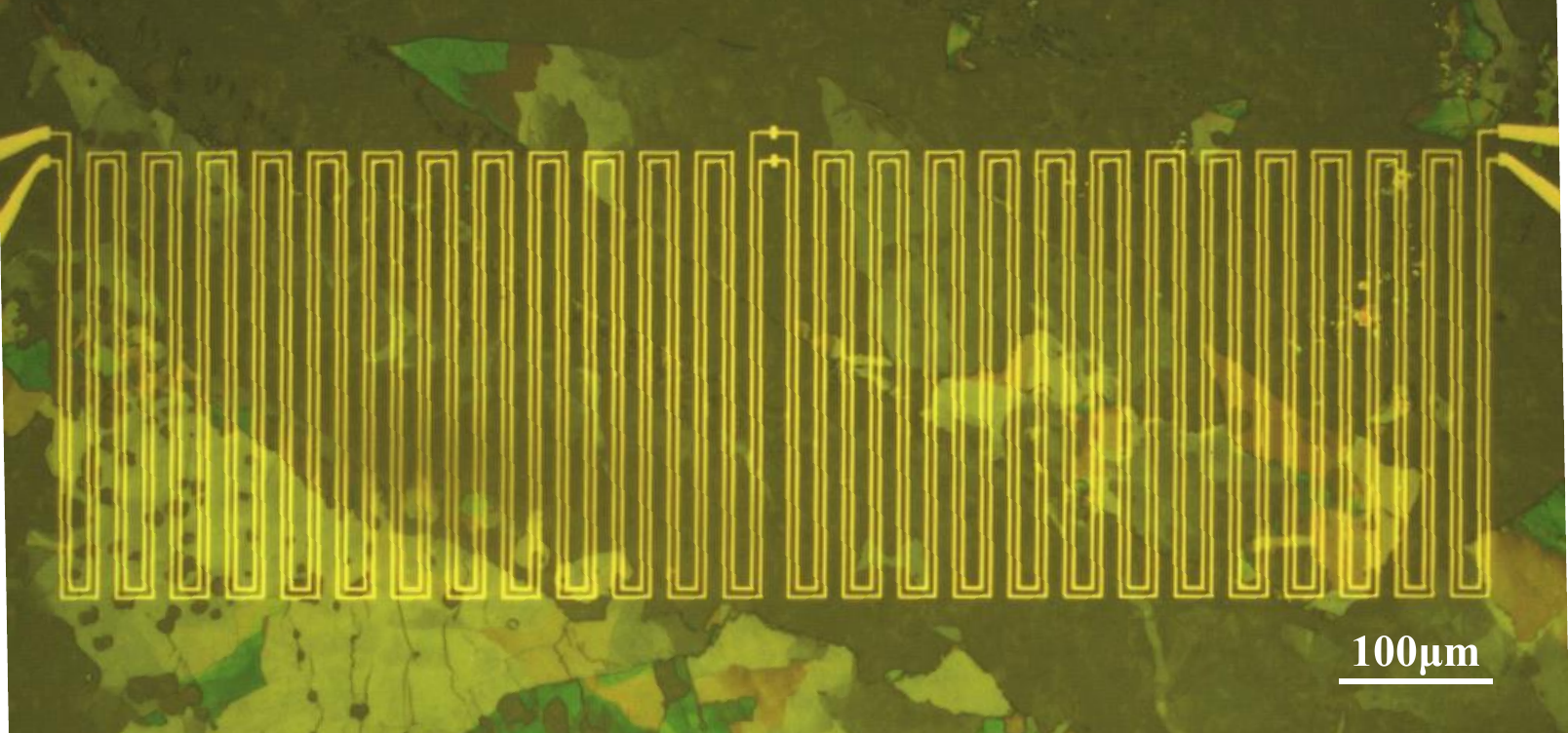}
    \caption[Optical micrograph of device]{Optical micrograph of mechanically exfoliated F-PEAI flakes transferred onto prefabricated interdigitated Au electrodes. Scale bar: 100~\si{\micro\metre}.}
    \label{fig:device}
\end{figure}

Figure~\ref{fig:device} shows mechanically exfoliated F-PEAI flakes transferred onto prefabricated Au electrodes. Thin flakes were obtained from bulk single crystals using thermal-release tape (Graphene Supermarket, SKU: GTT-5P) and inspected under a white-light optical microscope (Nikon LV150). Flakes exhibiting uniform optical contrast which indicate a uniform thickness and suitable lateral dimensions were selected for device fabrication. The transparency of the thermal-release tape enabled optical alignment of the selected flakes with the electrode pattern. Following alignment, the tape was brought into contact with the substrate and heated to 95~\si{\celsius} for 10~\si{\second}, releasing the adhesive and transferring the F-PEAI flakes onto the electrodes.

\clearpage{}
\section{Data}
\label{sec:data}

\begin{figure}[H]
    \centering
    \includegraphics[width=0.5\linewidth]{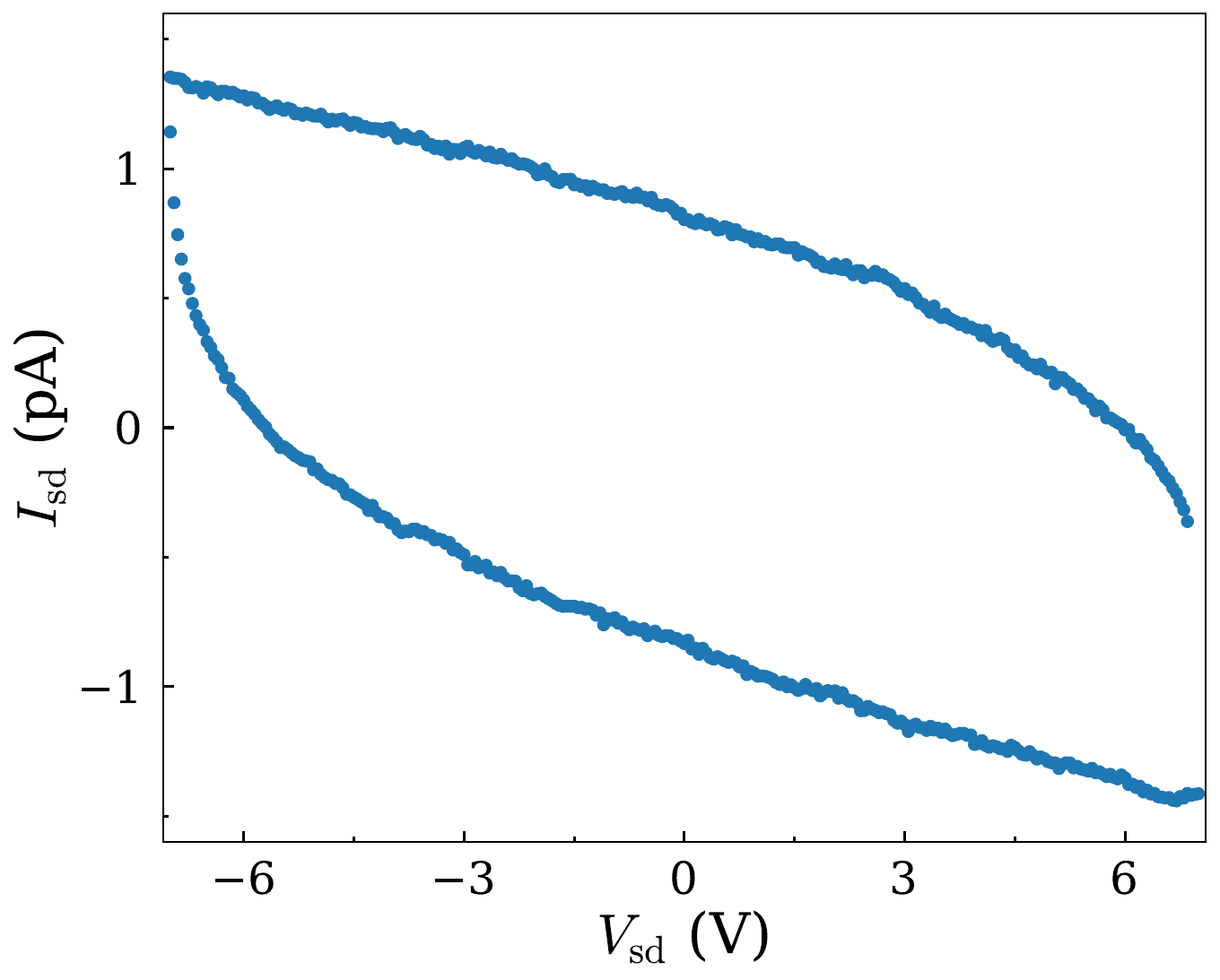}
    \caption[Dark current]{Device current measured in the dark for voltage for forward ($-7 \rightarrow +7$~\si{\volt}) and reverse voltage sweeps ($+7 \rightarrow -7$~\si{\volt}).}
    \label{fig:DarkCurrent}
\end{figure}

\begin{figure}[H]
\centering
\subfloat[Rejected sweep]{\includegraphics[height=0.3\linewidth]{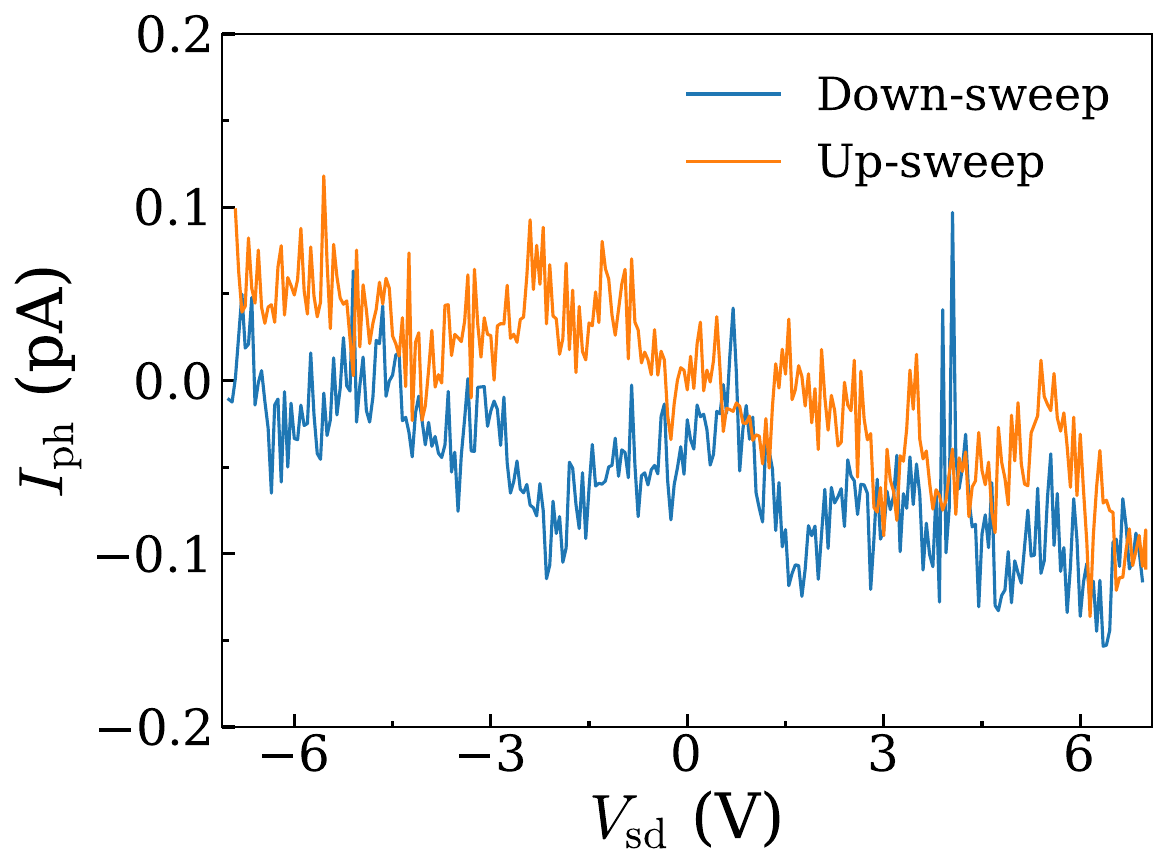}\label{fig:data_example:bad}}
\subfloat[Accepted sweep]{\includegraphics[height=0.3\linewidth]{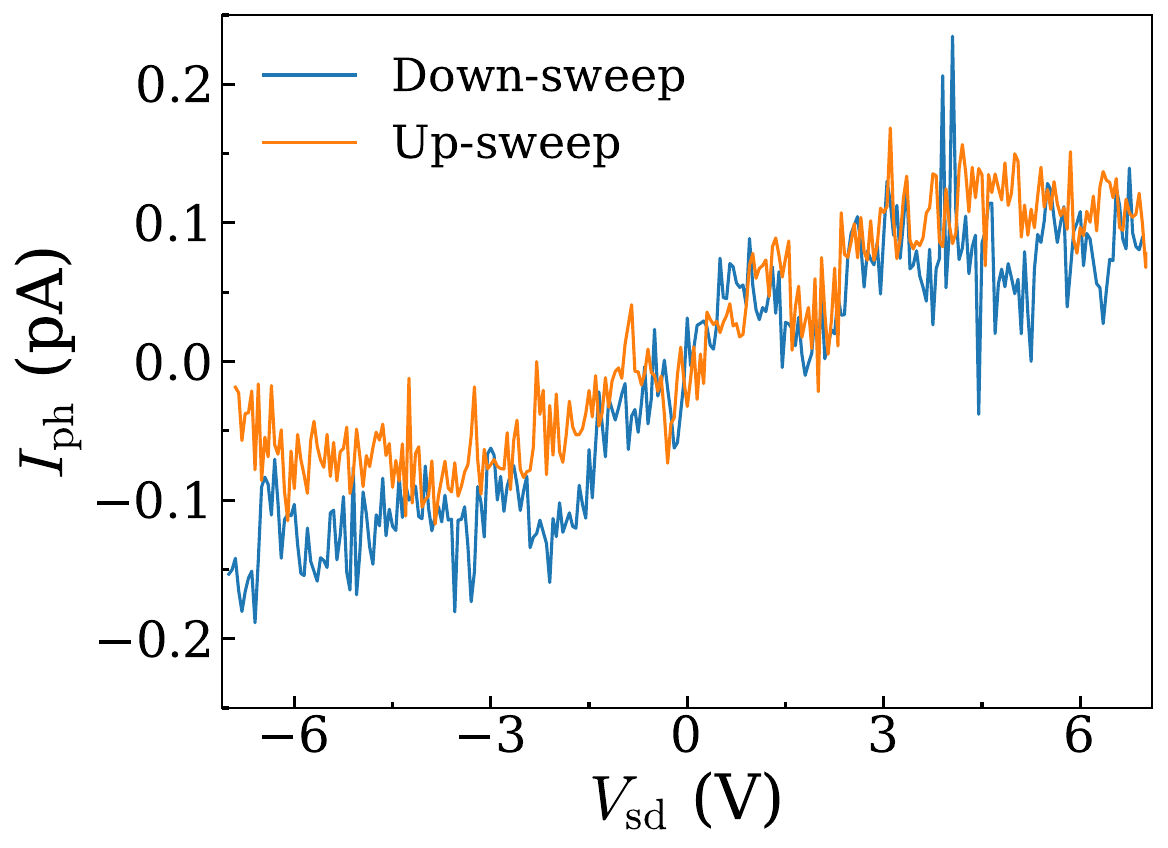}\label{fig:data_example:good}}
\caption[Photocurrent data examples]{Representative current-voltage sweeps illustrating the automated quality-control procedure. \protect\subref{fig:data_example:bad} Low-information sweep rejected before model training. \protect\subref{fig:data_example:good} Sweep retained for model development and evaluation.}
\label{fig:data_example}
\end{figure}

\textbf{Data quality control.}
The original experimental dataset comprised 505 photocurrent-voltage sweeps. Prior to machine-learning training, an automated quality-control procedure was applied to remove measurements exhibiting insufficient or noise-dominated bias-dependent modulation. For this screening, the source-drain voltage and photocurrent of each sweep were temporarily normalised by their respective maximum absolute values, and the forward and reverse branches were independently fitted with first-order polynomials. Sweeps for which either branch exhibited a normalised slope below 0.5, corresponding to less than a 50\% variation in normalised photocurrent across the applied bias range, were rejected. Representative rejected and accepted sweeps are shown in \figref{fig:data_example}.
\figurename~1e in the main text shows more examples of accepted sweeps in which signal-to-noise ratio is even higher.
This procedure removed 29 low-information measurements, leaving 476 sweeps for model development and evaluation.

\textbf{Model inputs and normalisation.}
The experimental dataset samples the nonlinear optoelectronic response of the photodetector across wavelength and irradiance.
The model inputs are the source-drain voltage ($V_\mathrm{sd}$), the photocurrent ($I_\mathrm{ph}$), and the measured current in the dark ($I_\mathrm{dark}$, i.e. the dark current).
The photocurrent is defined as

\begin{equation}
I_{\mathrm{ph}}(V_\mathrm{sd}) =
I_{\mathrm{sd}}(V_\mathrm{sd}) -
I_{\mathrm{dark}}(V_\mathrm{sd}) -
I_{\mathrm{offset}},
\end{equation}

\noindent
where $I_{\mathrm{sd}}$ is the measured current under illumination, and $I_\mathrm{offset}=0.0001$~\si{\nano\ampere} is a constant correction for instrumental current offsets.
Internally within the model, both current channels are normalised by the same scale factor, defined by the maximum absolute photocurrent within each voltage sweep;

\begin{equation}
I_{\mathrm{ph}}^\mathrm{norm}(V_\mathrm{sd}) =
\frac{
I_{\mathrm{ph}}(V_\mathrm{sd})
}{
\max_{V_\mathrm{sd}}
\left(
\left|
I_{\mathrm{ph}}(V_\mathrm{sd})
\right|
\right)
},
\end{equation}

\noindent
and

\begin{equation}
I_{\mathrm{dark}}^\mathrm{norm}(V_\mathrm{sd}) =
\frac{
I_{\mathrm{dark}}(V_\mathrm{sd})
}{
\max_{V_\mathrm{sd}}
\left(
\left|
I_{\mathrm{ph}}(V_\mathrm{sd})
\right|
\right)
}.
\end{equation}

\noindent
The absolute photocurrent magnitude is therefore removed, while the relative scale between the photocurrent and dark-current channels is retained. The dark-current channel captures the intrinsic bias-dependent device response in the absence of illumination, whereas the photocurrent channel describes the illumination-induced response.
Their relative magnitude and voltage dependence therefore retain information relevant to the incident irradiance.

The sampled values of $V_\mathrm{sd}$ are used to construct the Legendre-polynomial representation described in the main text, but are not passed point-by-point to the neural network. Only the physical sweep limits, $V_\mathrm{min}=-7$~\si{\volt} and $V_\mathrm{max}=+7$~\si{\volt}, are appended to the resulting Legendre coefficients as input features.

\begin{figure}[H]
\centering
\subfloat[]{\includegraphics[scale=0.46]{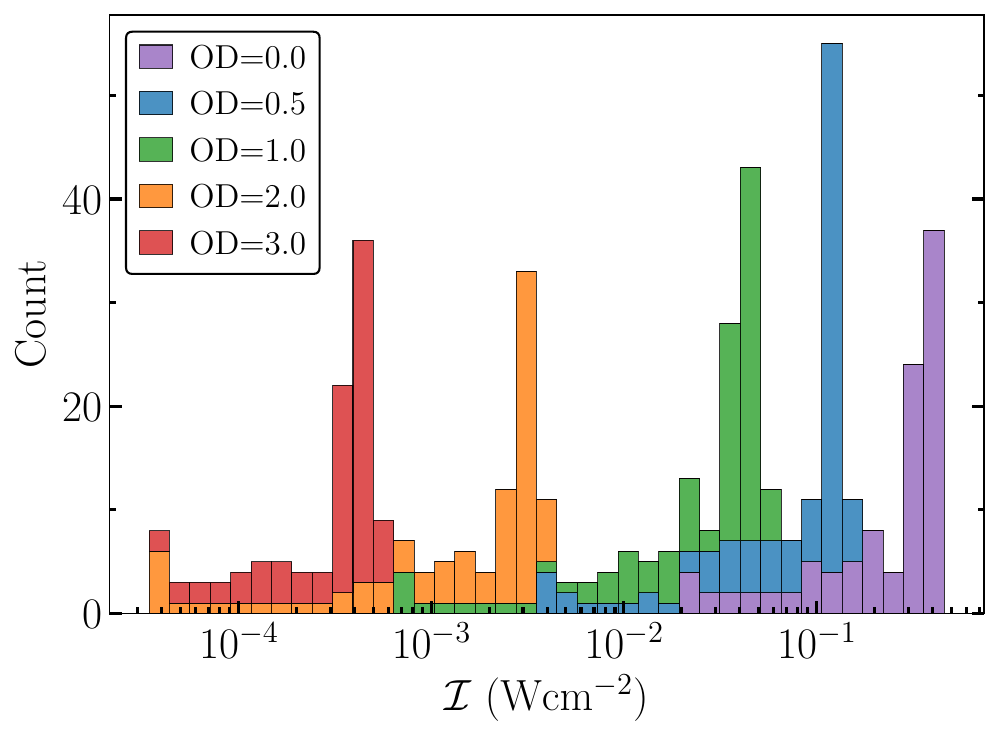}\label{fig:distribution:irradiance}}%
\hspace{1em}%
\subfloat[]{\includegraphics[scale=0.46]{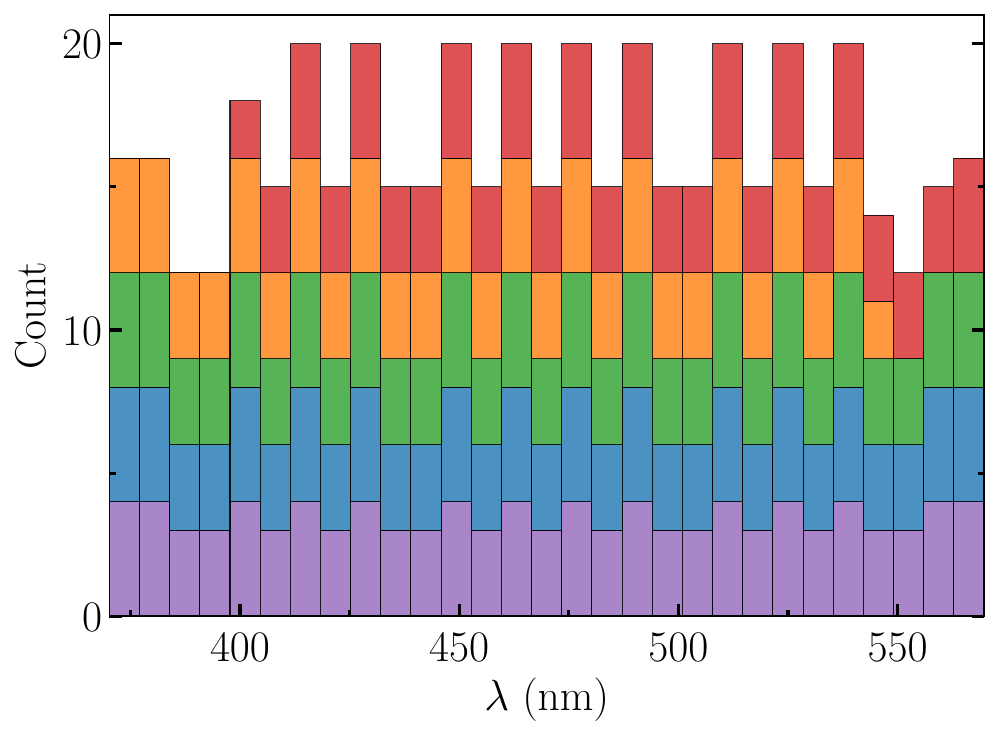}\label{fig:distribution:wavelength}}%
\caption[Data distribution]{Distribution of the accepted measurements across \protect\subref{fig:distribution:irradiance} incident irradiance and \protect\subref{fig:distribution:wavelength} excitation wavelength, resolved by optical-density filter.}
\label{fig:distribution}
\end{figure}

\textbf{Dataset coverage.}
The distribution of the 476 accepted measurements is shown in \figref{fig:distribution}. The excitation wavelengths are distributed nearly uniformly across the investigated spectral window from 370 to 570~\si{\nano\metre} (\figref{fig:distribution:wavelength}), limiting wavelength-dependent sampling bias during training. Under stronger attenuation, fewer usable measurements are retained near the spectral extremes because the reduced photocurrent leads to lower signal-to-noise ratios. Above approximately 520~\si{\nano\metre}, this reduction is further associated with the absorption edge of F-PEAI (\figurename~1c of the main text), beyond which the photocurrent decreases rapidly as the semiconductor absorption weakens. The resulting dataset therefore samples the spectral region in which the photodetector provides a measurable, information-rich nonlinear response.

The corresponding irradiance distribution is shown in \figref{fig:distribution:irradiance}. Increasing optical density systematically shifts the sampled irradiance towards lower values, producing an experimental range from $3.4\times10^{-5}$ to $4.6\times10^{-1}$~\si{\watt\per\centi\metre\squared}.
Adjacent optical-density-filter conditions exhibit substantial overlap in irradiance, with most spanning approximately two orders of magnitude. This overlap provides continuous coverage across attenuation conditions and ensures that comparable irradiance levels are represented under multiple experimental configurations, reducing the likelihood that the model learns isolated regions of the wavelength-irradiance parameter space.

\textbf{Generalisation and overfitting control.}
Despite the relatively small dataset of 476 accepted sweeps, several elements of the model and training procedure are designed to reduce overfitting.
First, the truncated Legendre-polynomial representation reduces the dimensionality of the measured current-voltage traces while suppressing high-order experimental noise.
Second, the variational latent representation introduces stochastic regularisation during training.
Third, self-supervised pretraining uses independently augmented views of the electrical response to learn a robust latent representation before supervised optimisation (\secref{sec:pretraining}).
Finally, excitation wavelength-irradiance pairs reserved for testing are explicitly excluded from the training set, ensuring that the reported wavelength reconstruction assesses generalisation to optical states not encountered during model optimisation.

\clearpage{}
\section{Self-supervised pretraining}
\label{sec:pretraining}

Before supervised optimisation, the encoder is pretrained using self-supervised learning (SSL) to learn robust representations of the unlabelled electrical responses. We employ the SimSiam framework~\cite{Chen2021ExploringSimpleSiamese}, which maximises agreement between two independently augmented views of the same input without requiring negative sample pairs. This provides a compact pretraining strategy that does not rely on large contrastive batches.

\textbf{Data augmentation.}
For each measured current-voltage response, two augmented views are generated by applying perturbations to the raw electrical traces before Legendre-polynomial decomposition.
The first (minor) augmentation view involves masking the zero-frequency component of the Fourier-transformed signal.
For second (major) augmentation view, the phase noise drawn from a zero-mean Gaussian distribution with a standard deviation of $0.1$ is introduced in the Fourier domain, followed by masking of the three lowest-frequency components.
These perturbations introduce controlled variability while retaining the dominant structure of the measured response, encouraging the encoder to learn representations that are insensitive to low-level experimental variations. The augmented traces are subsequently projected onto the same Legendre-polynomial basis used during supervised model training.

\textbf{SimSiam loss.}
Let $\mathbf{x}_1$ and $\mathbf{x}_2$ denote two augmented views of the same electrical response. Their encoded representations are projected through the SimSiam projection network to obtain $\mathbf{z}_1$ and $\mathbf{z}_2$, and subsequently passed through a prediction head to generate $\mathbf{p}_1$ and $\mathbf{p}_2$. The symmetric SimSiam objective is

\begin{equation}
\mathcal{L}_{\mathrm{simsiam}}
=
-\frac{1}{2}
\left[
\operatorname{cos}
\left(
\mathbf{p}_1,\operatorname{sg}(\mathbf{z}_2)
\right)
+
\operatorname{cos}
\left(
\mathbf{p}_2,\operatorname{sg}(\mathbf{z}_1)
\right)
\right],
\end{equation}

\noindent
where $\operatorname{sg}(\cdot)$ denotes the stop-gradient operation and $\operatorname{cos}(\cdot,\cdot)$ is the cosine similarity. The stop-gradient operation is essential to the SimSiam formulation, allowing each branch to predict the representation of the other while avoiding simultaneous optimisation of both targets towards a collapsed representation.

The projection network comprises two fully connected layers with ReLU activation and maps the encoder representation to a 64-dimensional space. The prediction head reduces this representation to 32 dimensions before projecting it back to 64 dimensions for evaluation of the SimSiam objective.

\textbf{KL-divergence regularisation.}
Because the encoder parameterises a variational latent distribution, pretraining additionally includes a Kullback-Leibler (KL) divergence term that regularises the approximate posterior towards a standard multivariate normal prior:

\begin{equation}
\mathcal{L}_{\mathrm{KL}}
=
D_{\mathrm{KL}}
\left[
\mathcal{N}
\left(
\boldsymbol{\mu},
\operatorname{diag}(\boldsymbol{\sigma}^{2})
\right)
\parallel
\mathcal{N}(\mathbf{0},\mathbf{I})
\right]
=
-\frac{1}{2}
\sum_{i=1}^{D_z}
\left(
1+\log\sigma_i^2-\mu_i^2-\sigma_i^2
\right),
\end{equation}

\noindent
where $\boldsymbol{\mu}$ and $\boldsymbol{\sigma}^2$ are the mean and variance predicted by the variational encoder and $D_z$ is the latent-space dimensionality. This regularisation promotes a smooth, continuous latent distribution and constrains the representation during self-supervised pretraining.

\textbf{Total pretraining objective.}
The complete SSL objective combines the SimSiam and KL-divergence terms,

\begin{equation}
\mathcal{L}_{\mathrm{SSL}}
=
\lambda_{\mathrm{simsiam}}
\mathcal{L}_{\mathrm{simsiam}}
+
\lambda_{\mathrm{KL}}
\mathcal{L}_{\mathrm{KL}}.
\end{equation}

\noindent
with $\lambda_{\mathrm{simsiam}}=0.75$ and $\lambda_{\mathrm{KL}}=1.0$. Pretraining is performed for 50 epochs using the AdamW optimiser with a learning rate of $1\times10^{-2}$, a 10-epoch linear warm-up and subsequent cosine-annealing schedule. The pretrained encoder weights are not frozen; they are used to initialise the subsequent supervised optimisation, during which the complete encoder-decoder is fine-tuned to map the electrical representation onto the Gaussian spectral parameters.

\clearpage{}
\section{Training loss function}
\label{sec:loss}

The supervised training objective is formulated directly in terms of the Gaussian components describing the incident spectrum. To support spectra containing multiple components, predicted and target peaks are associated through a differentiable soft-assignment procedure rather than a discrete matching operation. This preserves gradient flow through the matching step and enables end-to-end optimisation of the spectral decoder.

For each spectrum in a batch, the model predicts $K$ Gaussian components with centre wavelength $\mu_i$, amplitude $A_i$ and width $\sigma_i$, where $i\in\{1,\ldots,K\}$. The target spectrum contains $M$ components with corresponding parameters $\mu_j$, $A_j$ and $\sigma_j$, where $j\in\{1,\ldots,M\}$. In the present dataset all spectra are single-peaked, such that $K=M=1$; the more general formulation below is retained to support multi-component spectra.

The total supervised loss comprises contributions from peak position, amplitude, width and minimum-amplitude regularisation:

\begin{equation}
\mathcal{L}_{\mathrm{peak}}
=
\lambda_{\mathrm{pos}}\mathcal{L}_{\mathrm{pos}}
+
\lambda_{\mathrm{amp}}\mathcal{L}_{\mathrm{amp}}
+
\lambda_{\mathrm{sigma}}\mathcal{L}_{\mathrm{sigma}}
+
\lambda_{\mathrm{min\text{-}amp}}\mathcal{L}_{\mathrm{min\text{-}amp}},
\label{eq:loss}
\end{equation}

\noindent
with $\lambda_{\mathrm{pos}}=1.0$, $\lambda_{\mathrm{amp}}=1.2$, $\lambda_{\mathrm{sigma}}=0.2$ and $\lambda_{\mathrm{min\text{-}amp}}=0.3$ for the model reported in the main text.

\subparagraph{Soft assignment.}
Correspondence between predicted and target peaks is described by a soft-assignment matrix $\mathbf{W}\in\mathbb{R}^{K\times M}$. The assignments are obtained from the pairwise peak-position distances through a temperature-scaled softmax,

\begin{equation}
\mathbf{W}
=
\operatorname{softmax}
\left(
-\frac{\mathbf{D}}{\tau}
\right),
\end{equation}

\noindent
where $\mathbf{D}\in\mathbb{R}^{K\times M}$ has elements

\begin{equation}
D_{ij}
=
\left|
\mu_i-\mu_j
\right|,
\end{equation}

\noindent
and $\tau$ controls the sharpness of the assignment.
The softmax operation is evaluated over the second dimension (i.e. the across the target peaks, $M$).
During training, $\tau$ is annealed from $1.0$ to $0.1$, progressively sharpening the correspondence between predicted and target components.

The resulting soft target parameters associated with predicted component $i$ are

\begin{align}
\tilde{\mu}_i
&=
\sum_{j=1}^{M} W_{ij}\mu_j,
\\
\tilde{A}_i
&=
\sum_{j=1}^{M} W_{ij}A_j,
\\
\tilde{\sigma}_i
&=
\sum_{j=1}^{M} W_{ij}\sigma_j.
\end{align}

\noindent
The corresponding assignment strength is

\begin{equation}
s_i
=
\sum_{j=1}^{M}W_{ij},
\end{equation}

\noindent
which quantifies the association of predicted component $i$ with the target spectrum. Large $s_i$ identifies a strongly matched component, whereas weakly assigned components are treated as potentially spurious and suppressed through the amplitude loss.

\subparagraph{Position loss.}
Peak-position errors are penalised using the mean-squared deviation from the soft target positions,

\begin{equation}
\mathcal{L}_{\mathrm{pos}}
=
\frac{1}{BK}
\sum_{b=1}^{B}
\sum_{i=1}^{K}
\left(
\mu_i^{(b)}
-
\tilde{\mu}_i^{(b)}
\right)^2.
\end{equation}

\noindent
where $B$ is the batch size.

\subparagraph{Amplitude loss.}
The amplitude loss interpolates smoothly between matching an assigned target amplitude and suppressing an unassigned prediction:

\begin{equation}
\mathcal{L}_{\mathrm{amp}}
=
\frac{1}{BK}
\sum_{b=1}^{B}
\sum_{i=1}^{K}
\left[
(1-\alpha_i)
\left|
A_i^{(b)}
-
\tilde{A}_i^{(b)}
\right|
+
\alpha_i
\left|
A_i^{(b)}
\right|
\right].
\end{equation}

\noindent
where

\begin{equation}
\alpha_i
=
\operatorname{sigmoid}
\left[
10(0.5-s_i)
\right].
\end{equation}

\noindent
Strongly assigned components are therefore driven towards their target amplitudes, whereas weakly assigned components are driven towards zero.

To prevent collapse towards vanishing amplitudes when target peaks are present, an additional minimum-amplitude penalty is applied:

\begin{equation}
\mathcal{L}_{\mathrm{min\text{-}amp}}
=
\frac{1}{B}
\sum_{b=1}^{B}
\max
\left[
0,\,
0.5
\sum_{j=1}^{M}
A_j^{(b)}
-
\sum_{i=1}^{K}
A_i^{(b)}
\right].
\end{equation}

\noindent
This term requires the total predicted amplitude to reach at least half of the total target amplitude, providing a non-zero gradient when spectral components are excessively suppressed. For spectra containing no target peaks ($M=0$), an additional penalty drives all predicted amplitudes towards zero, suppressing false spectral components.

\subparagraph{Width loss.}
Peak widths are optimised using a smooth-L1 (Huber) loss weighted by the assignment strength,

\begin{equation}
\mathcal{L}_{\mathrm{sigma}}
=
\frac{1}{BK}
\sum_{b=1}^{B}
\sum_{i=1}^{K}
s_i\,
\ell_{\mathrm{smooth\text{-}L1}}
\left(
\sigma_i^{(b)},
\tilde{\sigma}_i^{(b)};
\beta=0.5
\right),
\end{equation}

\noindent
where $\beta=0.5$ is the Huber-loss threshold. Weighting by $s_i$ limits width optimisation to components associated with the target spectrum, preventing weakly assigned or spurious peaks from contributing appreciably to the width loss.

\section{Correlation Matrix Calculation}
\label{sec:correlation_calculation}

To quantify how the voltage-dependent photocurrent co-varies with the reconstructed optical quantities, correlation analyses are performed separately for the forward and reverse voltage sweeps using the absolute Pearson correlation coefficient~\cite{Pearson1895NoteRegressionInheritance}.

For each source--drain voltage $v_i$ (or binned voltage interval) and wavelength bin $\lambda_j$, we calculate the correlation between the photocurrent $I_n(v_i)$ and the reconstructed spectral intensity $S_n(\lambda_j)$ across the $N$ samples:

\begin{equation}
M^{\lambda}_{i,j}
=
\left|
\frac{
\displaystyle\sum_{n=1}^{N}
\left[I_n(v_i)-\bar{I}(v_i)\right]
\left[S_n(\lambda_j)-\bar{S}(\lambda_j)\right]
}{
\displaystyle
\sqrt{
\sum_{n=1}^{N}
\left[I_n(v_i)-\bar{I}(v_i)\right]^2
}
\sqrt{
\sum_{n=1}^{N}
\left[S_n(\lambda_j)-\bar{S}(\lambda_j)\right]^2
}
}
\right|.
\end{equation}

\noindent
Here, $\bar{I}(v_i)$ and $\bar{S}(\lambda_j)$ denote the corresponding sample means. The normalisation factors entering the standard deviations cancel in the Pearson coefficient. A value of zero is assigned when either variable has zero variance or when the correlation is undefined because of insufficient samples.

An analogous analysis is performed for the reconstructed irradiance $E$. Samples are grouped into irradiance bins, and for voltage point $v_i$ and irradiance bin $k$ we calculate

\begin{equation}
M^{E}_{i,k}
=
\left|
\operatorname{Corr}
\left(
I(v_i),E
\right)
\right|_{E\in\mathrm{bin}_k},
\end{equation}

\noindent
using only samples belonging to the corresponding irradiance bin. At least two samples are required within a bin for the correlation to be defined.

For visualisation of how the correlation is distributed across wavelength or irradiance at each voltage point, the matrices are normalised row-wise:

\begin{align}
\widetilde{M}^{\lambda}_{i,j}
&=
\frac{
M^{\lambda}_{i,j}
}{
\displaystyle\sum_{j'}M^{\lambda}_{i,j'}
},
\\
\widetilde{M}^{E}_{i,k}
&=
\frac{
M^{E}_{i,k}
}{
\displaystyle\sum_{k'}M^{E}_{i,k'}
}.
\end{align}

\noindent
This normalisation emphasises the relative distribution of correlation across spectral or irradiance bins for a given voltage point; it does not preserve the absolute correlation magnitude between different voltage points.

To quantify the voltage dependence of the reconstructed peak position directly, we additionally calculate the absolute Pearson correlation between $I(v_i)$ and the predicted peak wavelength

\begin{equation}
\lambda_{\mathrm{peak}}
=
\underset{\lambda}{\arg\max}\,S(\lambda),
\end{equation}

\noindent
together with the mean irradiance correlation across irradiance bins. These quantities identify the voltage regions in which the measured photocurrent co-varies most strongly with the reconstructed wavelength or irradiance. Because the analysis is based on correlation rather than gradient propagation or causal attribution, it should be interpreted as a measure of statistical association rather than model feature importance.

All calculations are performed independently for the forward sweep ($-7$~\si{\volt} to $+7$~\si{\volt}) and reverse sweep ($+7$~\si{\volt} to $-7$~\si{\volt}), preserving differences associated with hysteresis and sweep direction.

\clearpage{}
\section{Extended results}

\subsection{Irradiance parity}
\label{sec:results:irradiance}

\begin{figure}[H]
\centering
\includegraphics[scale=0.4]{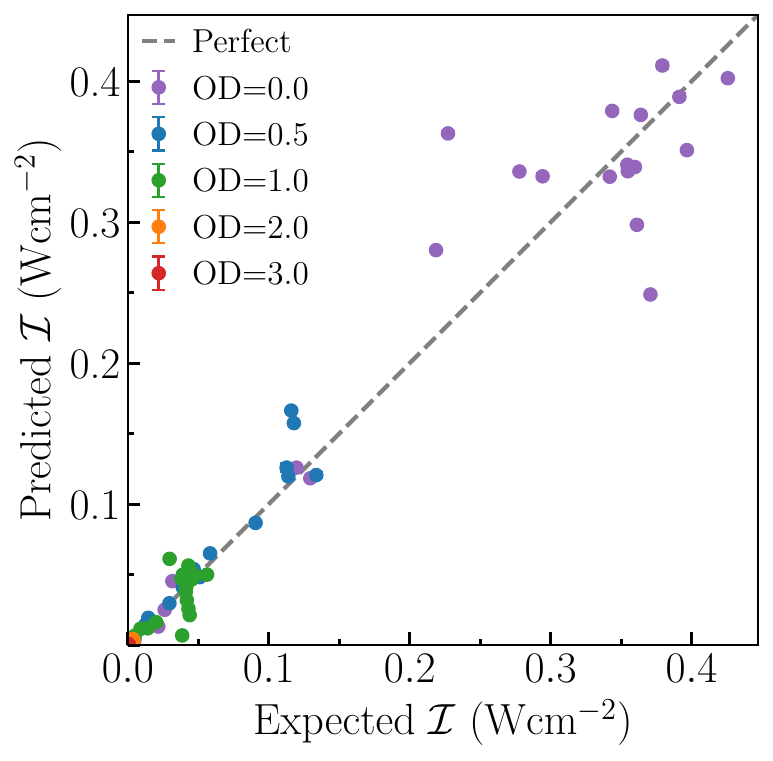}
\caption[Irradiance parity plot]{Parity between experimental and reconstructed irradiance for the held-out test set. The dashed line denotes ideal prediction, and the vertical whiskers show the spread of the 100-member latent-space ensemble.}
\label{fig:irradiance_parity}
\end{figure}

\Figref{fig:irradiance_parity} shows the reconstruction performance for irradiance expressed on its original linear scale, complementing the log-normalised representation reported in \figurename~3b of the main text. Because the experimental irradiance spans several orders of magnitude and the model is optimised in log-normalised space, the logarithmic representation provides a more appropriate measure of performance across the full dynamic range. On the linear irradiance scale, the ensemble reconstruction yields $R^2=0.9584\pm0.0170$, with a 95\% confidence interval of $[0.9228,\,0.9852]$.

\subsection{Deterministic versus ensemble inference}
\label{sec:results:deterministic}

The predictions reported in the main text are obtained from 100 samples of the learned latent distribution for each input measurement. The ensemble mean defines the reported prediction, while the spread across samples provides the model-internal sensitivity measure discussed in the main text. For comparison, deterministic inference can be performed using only the latent mean, $\mathbf{z}=\boldsymbol{\mu}$, without sampling from the learned variance. This produces a small increase in predictive accuracy but does not provide the corresponding ensemble spread.

For the same trained model and random seed used in the main text (seed 789), the current analysis gives deterministic $R^2$ values of 0.9588, 0.9871, and 0.9591 for wavelength, log-normalised irradiance, and irradiance, respectively.
Because the difference in predictive accuracy between deterministic and ensemble inference is small, while latent-space sampling additionally provides a measure of prediction sensitivity, ensemble-based results are reported throughout the main text.

\subsection{Optical-density filter-specific analysis}
\label{sec:results:od_filter}

\begin{table}[htbp]
\centering
\caption[OD filter-specific model metrics]{Wavelength-reconstruction performance resolved by optical-density (OD) filter.}
\label{tab:od_filter_performance}
\begin{tabular}{@{} c c c @{}}
\toprule
OD filter & Wavelength $R^2$ & Samples ($n$) \\
\midrule
0.0 & 0.9485 & 20 \\
0.5 & 0.9457 & 16 \\
1.0 & 0.9767 & 19 \\
2.0 & 0.9808 & 17 \\
3.0 & 0.9224 & 23 \\
\bottomrule
\end{tabular}
\end{table}

The $R^2$ metrics reported in the main text are evaluated across all test data, independent of OD filter; here, we break down the wavelength reconstruction performance for each filter individually.
This analysis uses the same trained model and test data as the main text (no new training or additional data are introduced).

\tabref{tab:od_filter_performance} resolves the model performance across the individual optical-density-filter conditions. Wavelength reconstruction remains consistently high across the investigated attenuation range, with $R^2=0.9224$--$0.9808$. The highest values are obtained for OD 1.0 and OD 2.0 ($R^2=0.9767$ and 0.9808, respectively), whereas the strongest attenuation, OD 3.0, gives the lowest value ($R^2=0.9224$), consistent with the reduced signal-to-noise ratio at low irradiance. The unattenuated and OD 0.5 measurements exhibit comparable performance ($R^2=0.9485$ and 0.9457, respectively).

Because each subgroup contains only 16--23 test measurements, these values are interpreted primarily as evidence that the reconstruction remains stable across attenuation conditions rather than as precise estimates of filter-specific performance.

\subsection{Correlation}
\label{sec:results:correlation}

\begin{figure}[H]
    \centering
    \subfloat[]{\includegraphics[width=0.48\linewidth]{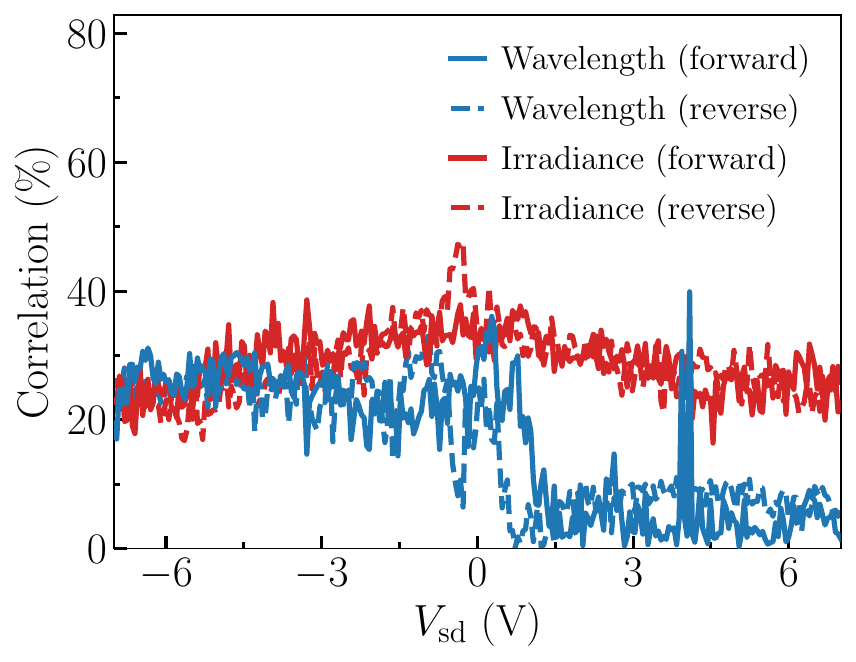}\label{fig:correlation:photocurrent}}%
    \hspace{1em}%
    \subfloat[]{\includegraphics[width=0.48\linewidth]{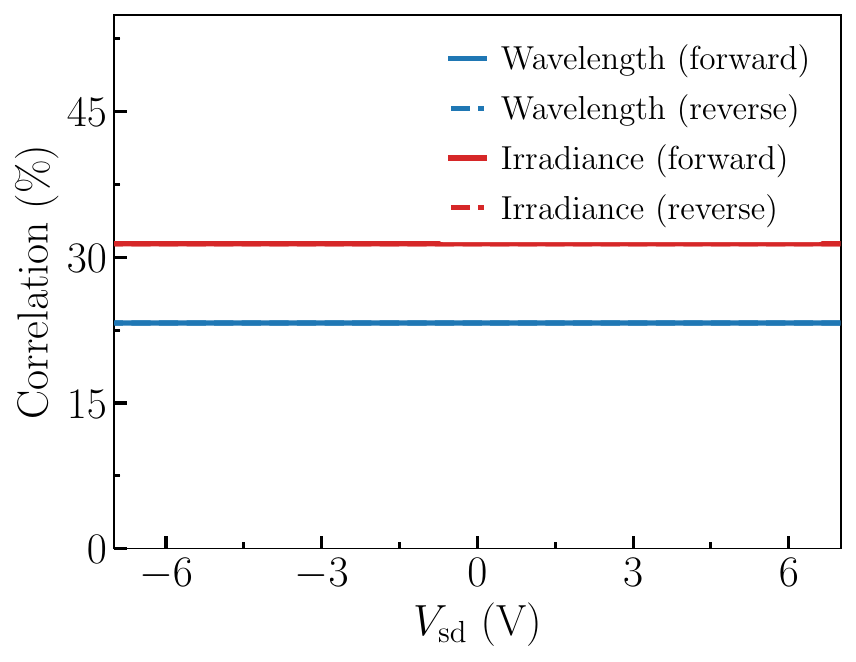}\label{fig:correlation:dark_current}}%
    \caption[Correlation between current and experimental spectral features]{Correlation between the \protect\subref{fig:correlation:photocurrent} photocurrent and \protect\subref{fig:correlation:dark_current} dark current at each source-drain voltage and experimental spectral features (wavelength and irradiance), summed over said spectral feature in the dataset.}
    \label{fig:correlation}
\end{figure}

\Figref{fig:correlation} shows the Pearson correlation coefficient between the (normalised) photocurrent and dark current (as defined in \secref{sec:data}) and the experimental wavelength and irradiance test targets, computed at each voltage point.

Because the dark current curve is identical across all test samples, its correlation with either target is expected to be constant across the voltage range.
Dark current exhibits a higher correlation with irradiance than with wavelength.
The forward and reverse bias sweeps exhibit the same correlation.

For photocurrent, the correlation with irradiance remains nearly constant across voltages (with a slightly higher correlation near zero bias), whereas the correlation with wavelength shows a clear bias toward negative voltages.
Since this is a wavelength-summed correlation, the large positive correlation observed at higher wavelengths for negative voltages dominates over the contributions from other wavelength regions.
Two peaks in the correlation on wavelength are seen at around 4~\si{\volt}; this is attributed to the noise visible in the bad (and to a less extent, the good) data plotted in \figref{fig:data_example} at the same values (also present in \figurename.~5 of the main text).
As this is a Pearson correlation, rather than direct gradient propagation, these plots only present correlation, not causation, so noise in the input data will still show some correlation where there wouldn't exist any causation in the model.

\begin{figure}
    \centering
    \subfloat[\vspace{-0.75em}]{\includegraphics[scale=0.425]{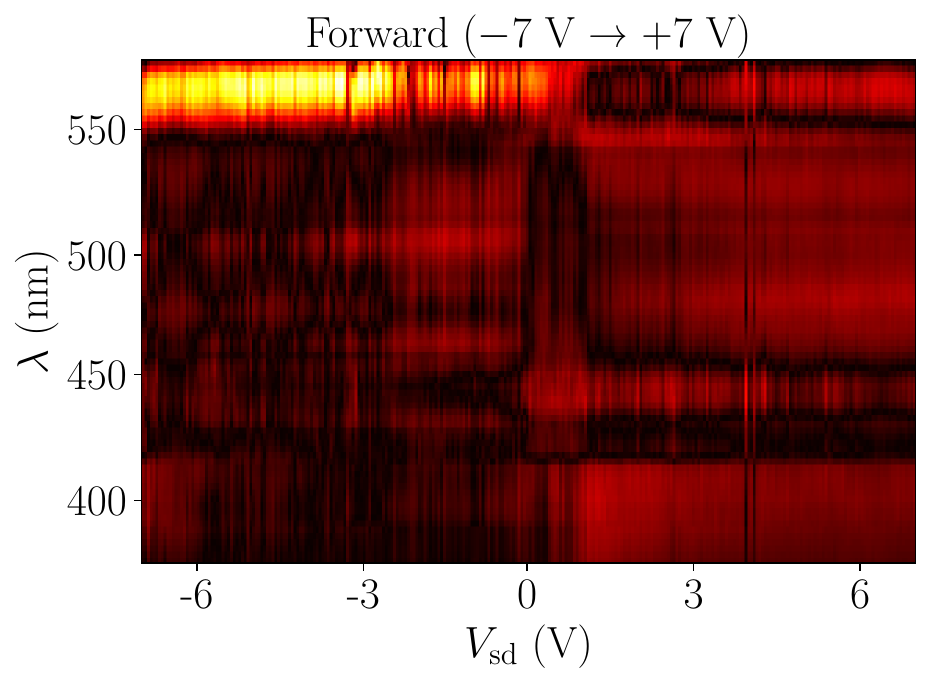}\label{fig:reconstructed_correlation:forward}}%
    \hspace{0.5em}%
    \subfloat[\vspace{-0.75em}]{\includegraphics[scale=0.425]{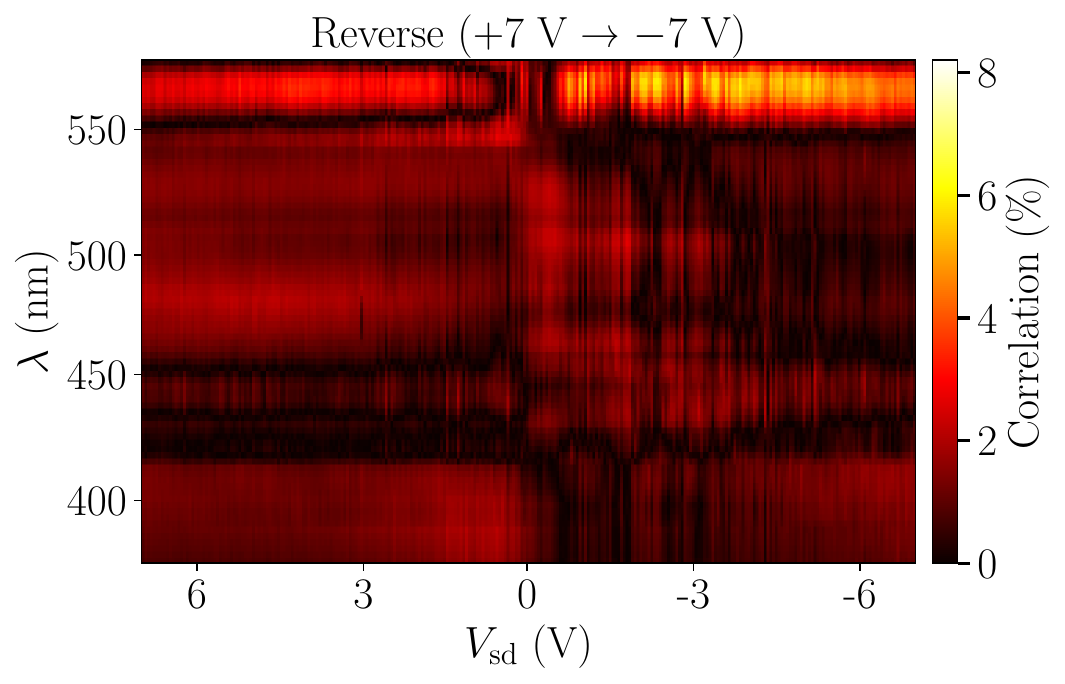}\label{fig:reconstructed_correlation:reverse}}%
    \vspace{-1em}\hspace{0.5em}%
    \subfloat[\vspace{-0.75em}]{\includegraphics[scale=0.425]{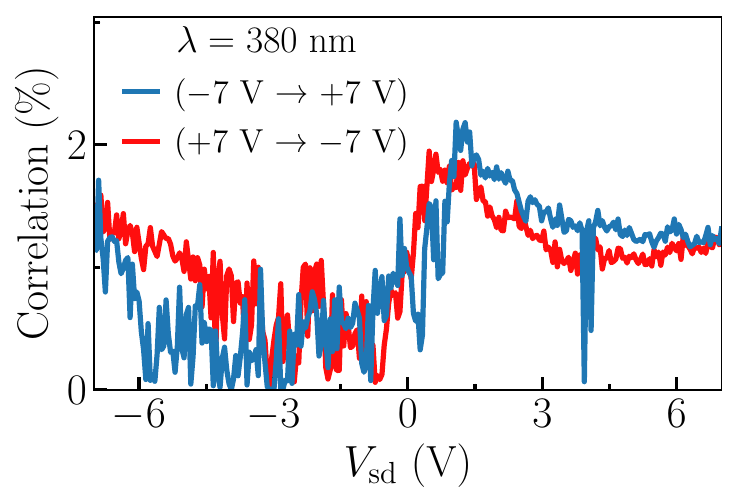}\label{fig:reconstructed_correlation:380}}%
    \hspace{0.5em}%
    \subfloat[\vspace{-0.75em}]{\includegraphics[scale=0.425]{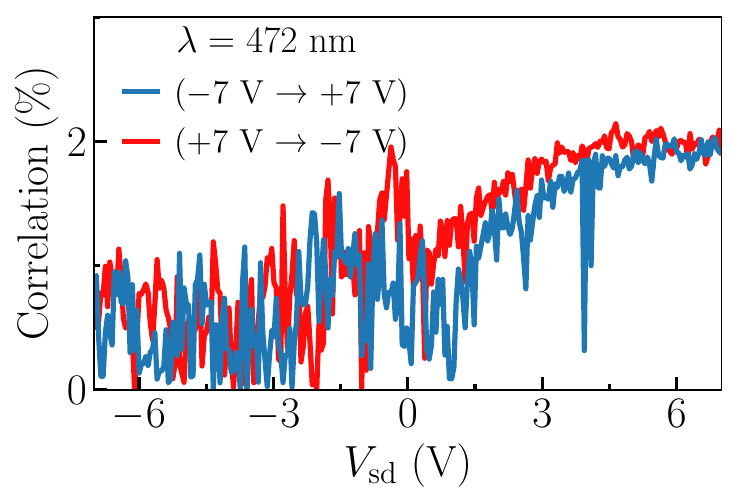}\label{fig:reconstructed_correlation:472}}%
    \hspace{0.5em}%
    \subfloat[\vspace{-0.75em}]{\includegraphics[scale=0.425]{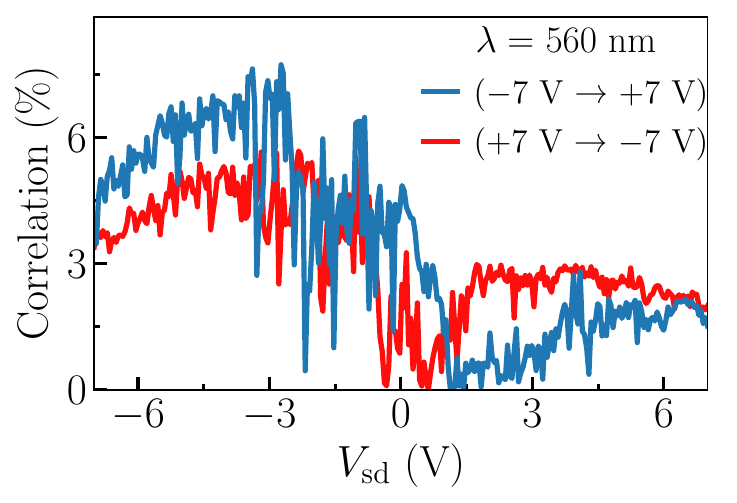}\label{fig:reconstructed_correlation:560}}%
    \caption[Correlation between current and reconstructed spectral features]{%
        Row-normalised absolute Pearson correlation between photocurrent and reconstructed spectral wavelength as a function of source-drain voltage for the \protect\subref{fig:reconstructed_correlation:forward} forward and \protect\subref{fig:reconstructed_correlation:reverse} reverse bias sweeps.
        Representative voltage-dependent correlation profiles at \protect\subref{fig:reconstructed_correlation:380} 380, \protect\subref{fig:reconstructed_correlation:472} 472, and \protect\subref{fig:reconstructed_correlation:560} 560~\si{\nano\metre}, comparing the forward and reverse sweep directions.
        The reconstructed spectra consist of the 95 samples in the test set.%
    }
    \label{fig:reconstructed_correlation}
\end{figure}

\Figref{fig:reconstructed_correlation} presents the correlation between photocurrent and reconstructed wavelength as a function of source-drain voltage (this figure is the equivalent of \figurename~5 from the main text, where the main text figure presents this correlation as a function of experimental wavelength).

\clearpage{}
\section{Robustness Analysis}
\label{sec:robustness}

\subsection{Training dataset size}
\label{sec:robustness:dataset_size}

\begin{figure}[H]
\centering
\includegraphics[width=0.65\linewidth]{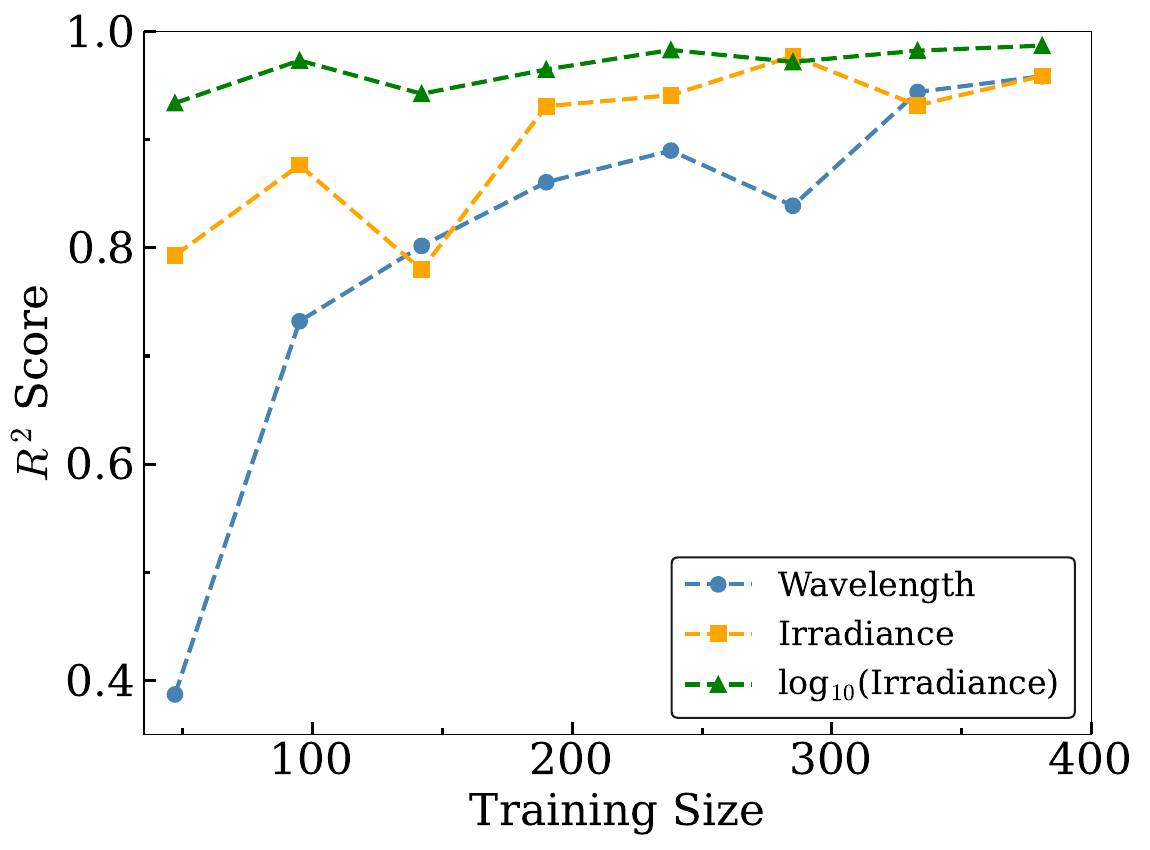}
\caption[Model metrics versus training data size]{$R^2$ as a function of training-set size for wavelength, irradiance and log-normalised irradiance reconstruction. The same held-out test set of 95 measurements is used for all comparisons.}
\label{fig:dataset_size}
\end{figure}

To assess the dependence of predictive performance on the amount of training data, the model was trained using 10, 20, 30, 40, 50, 60, 70 and 80\% of the available dataset, corresponding to 47--381 current--voltage sweeps. The held-out test set was fixed at the same 95 measurements for all comparisons. The resulting $R^2$ values are shown in \figref{fig:dataset_size}.
The metrics presented here are the deterministic $R^2$ metrics, i.e. the ensemble approach to statistical sampling of the latent space is not utilised here.

Performance improves overall as the training-set size increases, although finite-sample fluctuations produce some non-monotonic behaviour. Log-normalised irradiance is reconstructed accurately even with comparatively small training sets, whereas wavelength reconstruction exceeds $R^2=0.9$ only when approximately 70\% of the available data (333 sweeps) are used for training. Reconstruction of irradiance on the linear scale exhibits greater variability because the experimental values span several orders of magnitude, causing errors at the upper end of the irradiance range to contribute disproportionately to $R^2$. None of the three metrics exhibits a clear performance plateau at the maximum training size, indicating that reconstruction accuracy is not yet saturated by the available dataset and may improve with additional experimental measurements.

\subsection{Random-seed and data-split variability}
\label{sec:robustness:random_seed}

\begin{figure}[H]
\centering
\includegraphics[width=0.65\linewidth]{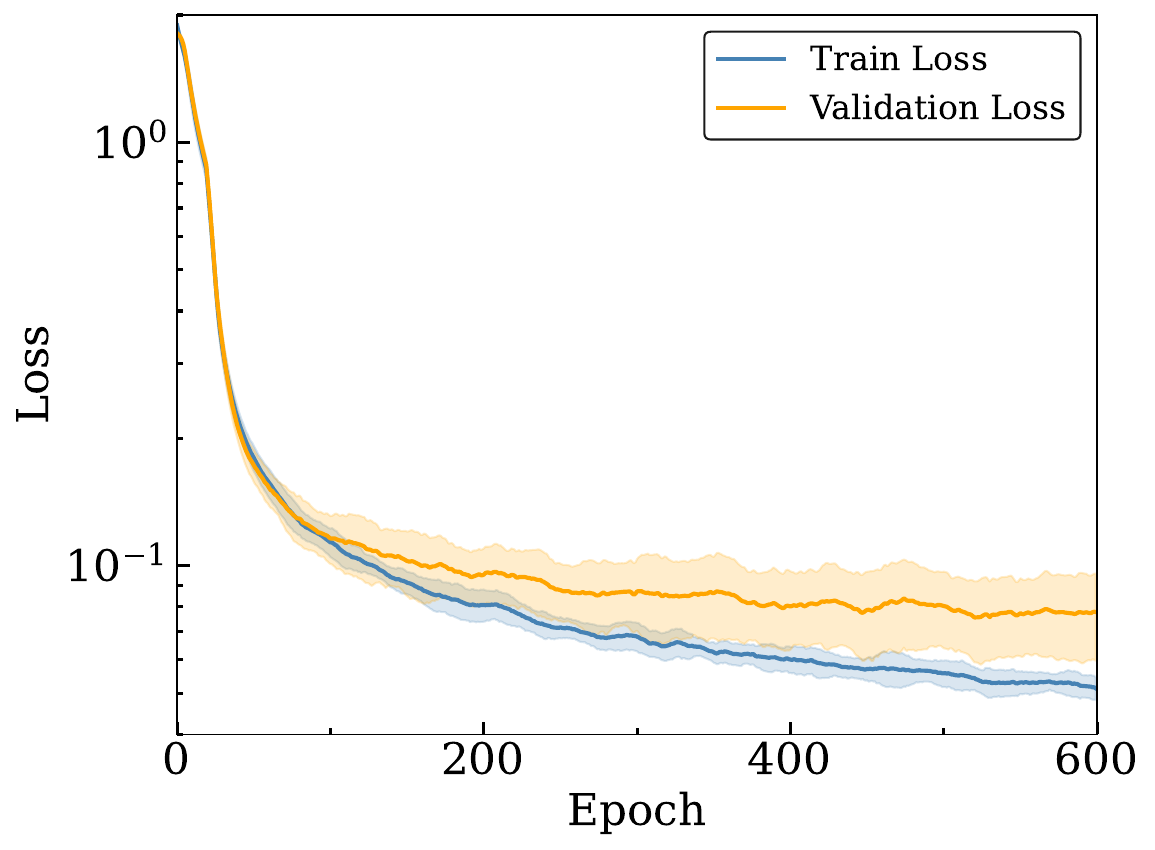}
\caption[Model training curves]{Training and validation loss across 20 independent random seeds and data splits. Solid curves show the mean loss and shaded regions denote one standard deviation across the 20 runs. A 20-epoch running average is applied for visualisation.}
\label{fig:training}
\end{figure}

Robustness to stochastic model initialisation and train/test partitioning was evaluated using 20 random seeds ($\{$0, 7, 33, 42, 256, 271, 314, 482, 789, 1618, 2024, 7617, 8030, 9041, 9876, 1049, 12345, 22018, 110011, 305769$\}$). For each seed, the model was trained from scratch using the same hyperparameters and an 80\% training fraction, while both the model initialisation and dataset partition were varied.
As with the model presented in the main text, wavelength-irradiance pairs assigned to the test set are explicitly excluded from the corresponding training set.
The metrics presented here are the deterministic $R^2$ metrics, i.e. the ensemble approach to statistical sampling of the latent space is not utilised here.

Figure~\ref{fig:training} shows the mean training and validation losses across the 20 runs. The validation loss approaches a plateau by approximately 600 epochs, while the training loss continues to decrease gradually. The limited spread between independent runs indicates stable optimisation across model initialisations and dataset partitions.

\begin{table}[htbp]
\centering
\caption[Statistical model metrics]{Summary statistics for $R^2$ across 20 random seeds (used for model initialisation and data partitioning). Metrics presented here are evaluated using deterministic inference. The 95\% confidence intervals were estimated by bootstrap resampling~\cite{Efron1979BootstrapMethodsAnother}.}
\label{tab:r2_stats}
\begin{tabular}{@{} l c c c c c @{}}
\toprule
Metric & Mean & Std. Dev. & Min & Max & 95\% CI \\
\midrule
Wavelength $R^2$ & 0.9463 & 0.0115 & 0.9255 & 0.9665 & [0.9412, 0.9513] \\
Log-normalised irradiance $R^2$ & 0.9840 & 0.0032 & 0.9782 & 0.9882 & [0.9826, 0.9853] \\
Irradiance $R^2$ & 0.9370 & 0.0264 & 0.8435 & 0.9659 & [0.9238, 0.9474] \\
\bottomrule
\end{tabular}
\end{table}

The corresponding summary statistics are reported in \tabref{tab:r2_stats}. Wavelength reconstruction yields a mean $R^2=0.9463$ with a standard deviation of 0.0115, while log-normalised irradiance is particularly stable, with $R^2=0.9840\pm0.0032$. Irradiance evaluated on the linear scale shows greater variability ($R^2=0.9370\pm0.0264$), consistent with its broad dynamic range. The narrow bootstrap confidence intervals obtained from 10,000 resampling iterations further demonstrate that the predictive performance is reproducible across stochastic initialisation and dataset partitioning.

\subsection{Omitting wavelengths from training data}
\label{sec:robustness:wavelengths}

\begin{figure}
    \centering%
    \subfloat[\vspace{-0.75em}]{\includegraphics[scale=0.3775]{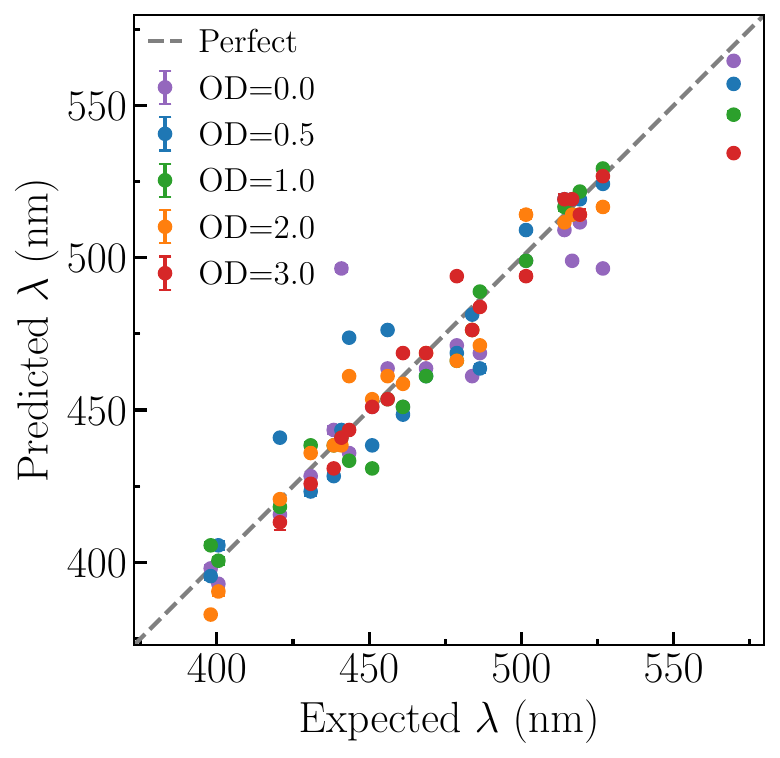}\label{fig:parity_unique:wavelength}}%
    \hspace{0.5em}%
    \subfloat[\vspace{-0.75em}]{\includegraphics[scale=0.3775]{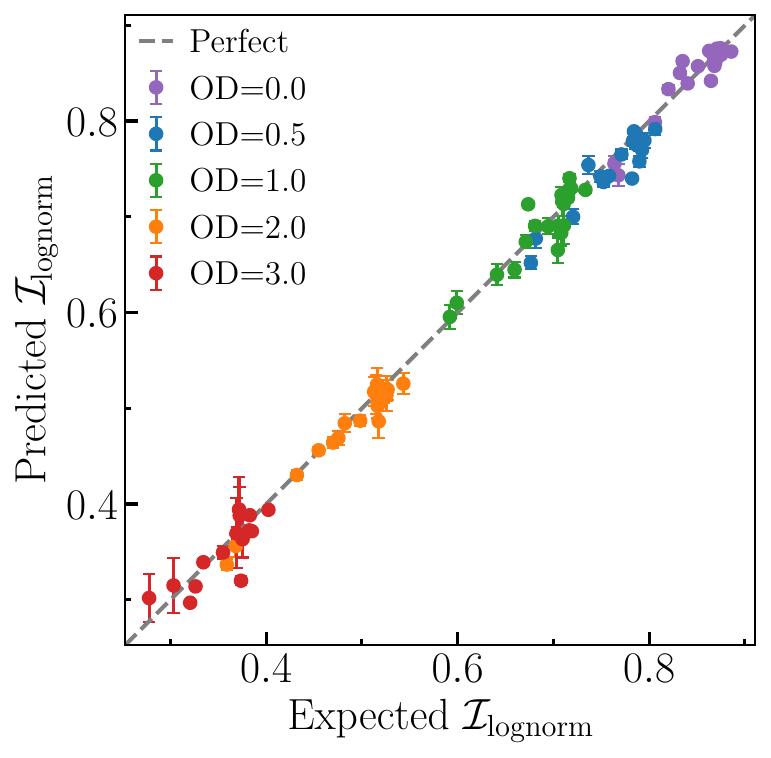}\label{fig:parity_unique:log_irradiance}}%
    \hspace{0.5em}%
    \subfloat[\vspace{-0.75em}]{
    \includegraphics[scale=0.3775]{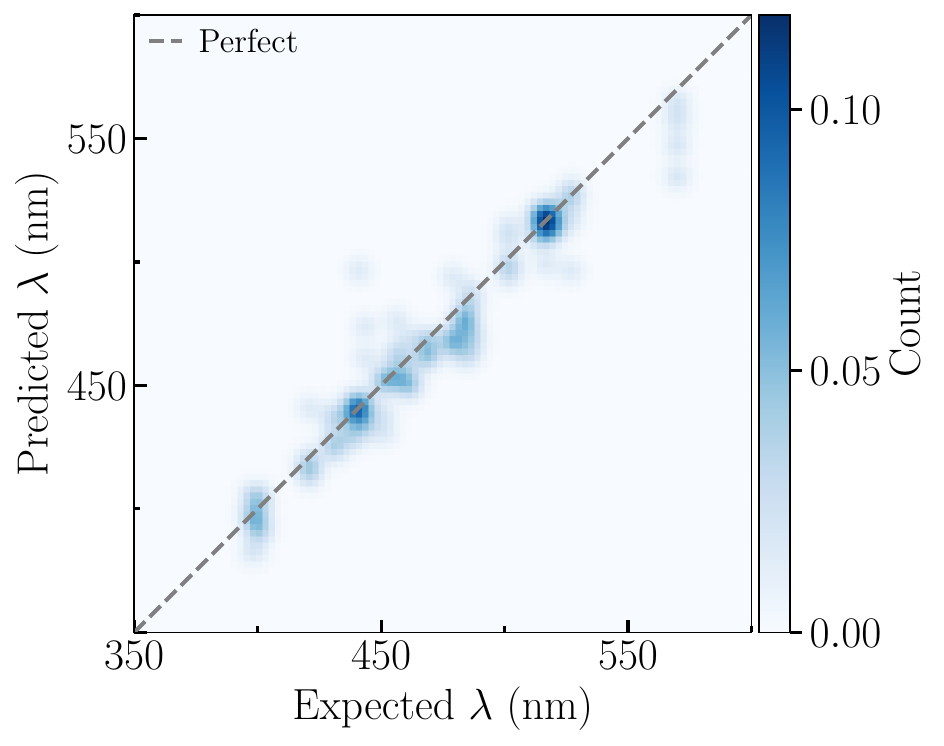}\label{fig:parity_unique:wavelength_confusion}}
    \caption[Predictive performance of the model]{%
        \textbf{Predictive performance of the spectral reconstruction model for omitted wavelengths.}
        Reconstructed versus experimental \protect\subref{fig:parity_unique:wavelength} wavelength and \protect\subref{fig:parity_unique:log_irradiance} log-normalised irradiance for the omitted wavelength held-out test set.
        The dashed line denotes ideal prediction; colours indicate optical-density filter (OD) level and vertical whiskers show the spread of the latent-space ensemble.
        \protect\subref{fig:parity_unique:wavelength_confusion} Prediction-density map of reconstructed versus experimental wavelength, highlighting the wavelength dependence of the reconstruction accuracy.%
    }
    \label{fig:parity_unique}
\end{figure}

Here we test the model's ability to predict wavelengths not represented in the training set, omitting entire wavelengths rather than individual wavelength-irradiance pairs.
The 101 unique wavelength bins were randomly partitioned into 81 training and 20 test wavelengths (yielding 379 training and 97 test samples), with the same model architecture, training procedure, and 20 random seeds as the other robustness experiments (see \secref{sec:robustness:random_seed} for the set of random seeds).
Unless stated otherwise, the metrics presented here are the deterministic $R^2$ metrics, i.e. the ensemble approach to statistical sampling of the latent space is not utilised here.

Despite the complete absence of test wavelengths from training, the model maintains strong performance: wavelength $R^2=0.9191$ (range $0.9013$--$0.9305$), log irradiance $R^2=0.9936$ (range $0.9905$--$0.9954$), and linear irradiance $R^2=0.9545$ (range $0.9086$--$0.9793$).
The slight degradation in wavelength reconstruction relative to the reference model ($R^2=0.958$) indicates that the architecture still generalises beyond discrete training wavelengths rather than simply memorising them.
Irradiance remains effectively unchanged, supporting the conclusion that it is encoded through broad, distributed features of the response rather than wavelength-specific bins.

The equivalent parity plots from \figurename~3 of main text are provided in \figref{fig:parity_unique}, where the seed closest to the statistical mean $R^2$ is used (seed $=1049$, deterministic wavelength $R^2=0.9198$, log irradiance $R^2=0.9926$).
Using the ensemble inference approach, the model achieves wavelength $R^2=0.9181\pm0.0239$ (95\% CI $0.8650$--$0.9541$), log irradiance $R^2=0.9924\pm0.0015$ (95\% CI $0.9893$--$0.9950$), and linear irradiance $R^2=0.9640\pm0.0105$ (95\% CI $0.9402$--$0.9815$), with mean uncertainties of $0.76$~\si{\nano\metre}, $7\times10^{-4}$~\si{\watt\per\centi\metre\squared}, and $0.0079$ for wavelength, irradiance, and log irradiance, respectively.
Interestingly, whilst the $R^2$ has decreased, the MAE and RMSE have changed very little (MAE $=8.1$~\si{\nano\metre} and RMSE $= 11.9$~\si{\nano\metre}).
When evaluating the $R^2$ metrics for each optical density filter separately, the reduction in overall wavelength performance is clearly attributable to the unattenuated condition (OD 0.0, $R^2=0.8522$).
All attenuated filters perform substantially better: OD 0.5 ($R^2=0.9170$), OD 1.0 ($R^2=0.9592$), OD 2.0 ($R^2=0.9479$), and OD 3.0 ($R^2=0.9315$); the subgroup sizes are $n=18$--$20$, showing an even split over OD filters for the test set.

These results demonstrate that the model can generalise to unseen wavelengths with only a $\sim0.03$ drop in $R^2$ and no drop in mean absolute error.

\subsection{Latent-space sampling sensitivity}
\label{sec:robustness:latent_space}

For the model reported in the main text, the variational encoder learns a mean latent standard deviation of approximately 0.04, with individual values spanning 0.007--0.700. To assess the sensitivity of the reconstructed spectrum to latent perturbations, ensemble inference was repeated using a fixed sampling scale of 0.5 in place of the learned standard deviation. This imposed perturbation is more than an order of magnitude larger than the mean learned scale.

Despite this substantially increased latent perturbation, predictive performance changes only weakly. As summarised in \tabref{tab:noise_comparison}, wavelength $R^2$ decreases from $0.9584\pm0.0103$ to $0.9560\pm0.0109$, irradiance $R^2$ from $0.9584\pm0.0170$ to $0.9525\pm0.0197$, and log-normalised irradiance $R^2$ from $0.9870\pm0.0048$ to $0.9867\pm0.0048$. The corresponding confidence intervals overlap strongly, indicating that the degradation in predictive accuracy is small relative to the variability of the ensemble.

\begin{table}[htbp]
\centering
\caption[Model metrics versus latent space noise]{Comparison of ensemble prediction metrics obtained using a fixed latent sampling scale of 0.5 and the learned latent standard deviation for the held-out test set ($n=95$).}
\label{tab:noise_comparison}
\begin{tabular}{@{}lcc@{}}
\toprule
\textbf{Metric} & \textbf{Fixed scale (0.5)} & \textbf{Learned std.} \\
\midrule
\multicolumn{3}{c}{\textbf{Performance metrics}} \\
\midrule
Wavelength $R^2$ & $0.9560 \pm 0.0109$ & $0.9584 \pm 0.0103$ \\
Wavelength $R^2$ 95\% CI & $[0.9321, 0.9742]$ & $[0.9360, 0.9759]$ \\
Irradiance $R^2$ & $0.9525 \pm 0.0197$ & $0.9584 \pm 0.0170$ \\
Irradiance $R^2$ 95\% CI & $[0.9110, 0.9831]$ & $[0.9228, 0.9852]$ \\
Log-normalised irradiance $R^2$ & $0.9867 \pm 0.0048$ & $0.9870 \pm 0.0048$ \\
Log-normalised irradiance $R^2$ 95\% CI & $[0.9753, 0.9933]$ & $[0.9755, 0.9936]$ \\
\midrule
\multicolumn{3}{c}{\textbf{Latent-sampling ensemble spread}} \\
\midrule
Mean wavelength spread (\si{\nano\metre}) & $5.9195$ & $0.6453$ \\
Mean irradiance spread (\si{\watt\per\centi\metre\squared}) & $2\times10^{-2}$ & $7\times10^{-4}$ \\
Mean log-normalised irradiance spread & $0.0865$ & $0.0065$ \\
\bottomrule
\end{tabular}
\end{table}

The principal effect of increasing the latent sampling scale is therefore not a large change in predictive accuracy, but a substantial increase in the spread of the ensemble. The mean wavelength spread increases from 0.65~\si{\nano\metre} to 5.92~\si{\nano\metre}, while the log-normalised irradiance spread increases from 0.0065 to 0.0865. The larger wavelength spread approaches the empirical wavelength MAE of approximately 8~\si{\nano\metre}; however, this numerical similarity should not be interpreted as evidence of uncertainty calibration. The ensemble spread quantifies sensitivity to perturbations of the learned latent representation and does not account for all sources of predictive error.

These results instead demonstrate that the inferred wavelength and irradiance remain stable under latent perturbations substantially larger than the mean learned sampling scale, while the ensemble spread responds strongly to the imposed perturbation. This supports the use of latent-space sampling as a model-internal sensitivity measure and confirms that the headline predictive performance is not strongly dependent on the precise magnitude of the learned latent variance.

\begin{figure}[H]
\centering
\subfloat[]{%
\includegraphics[scale=0.4]{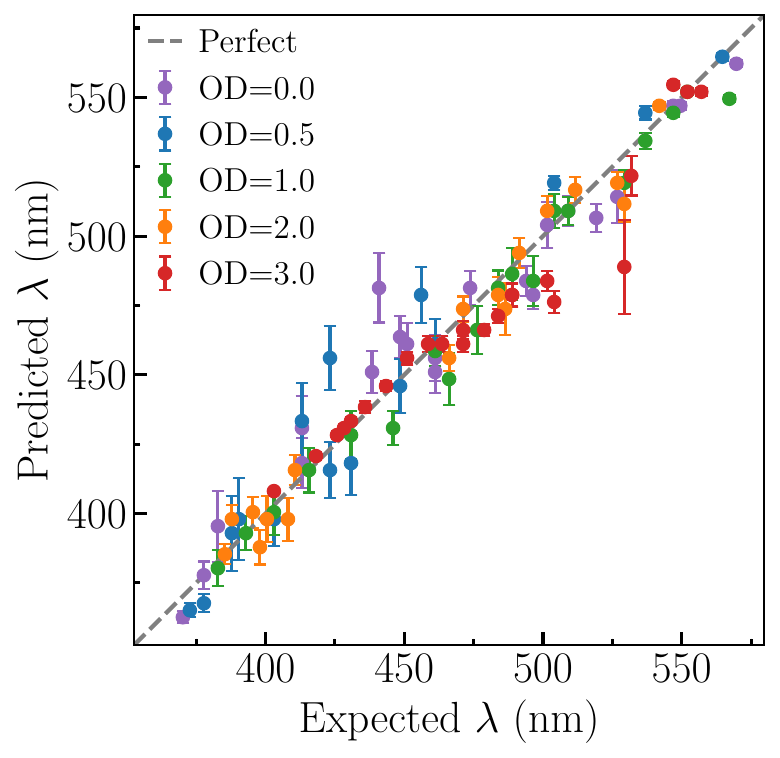}%
\label{fig:parity:wavelength}%
}%
\hspace{1em}%
\subfloat[]{%
\includegraphics[scale=0.4]{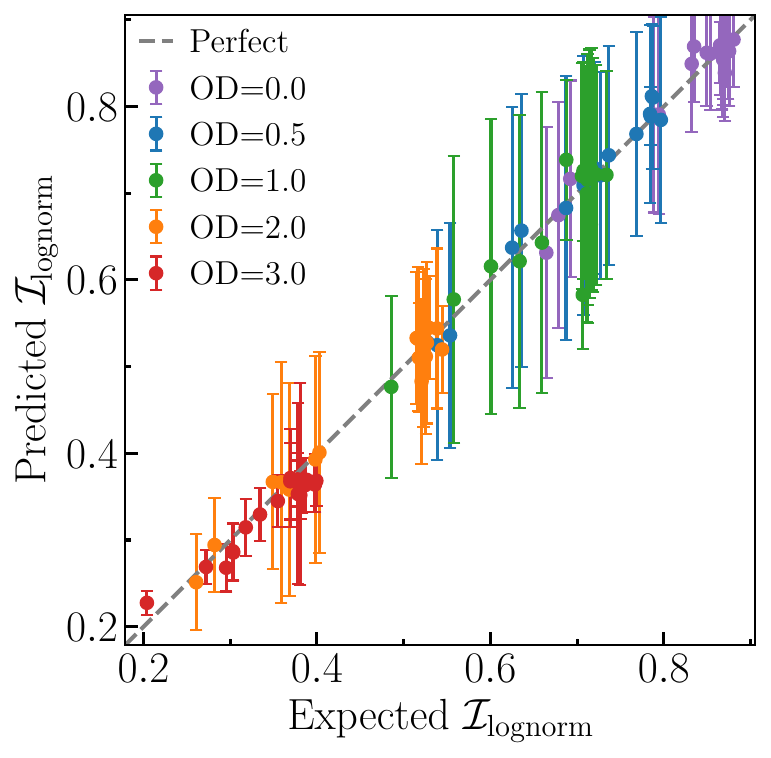}%
\label{fig:parity:log_irradiance}%
}%
\hspace{1em}%
\subfloat[]{%
\includegraphics[scale=0.4]{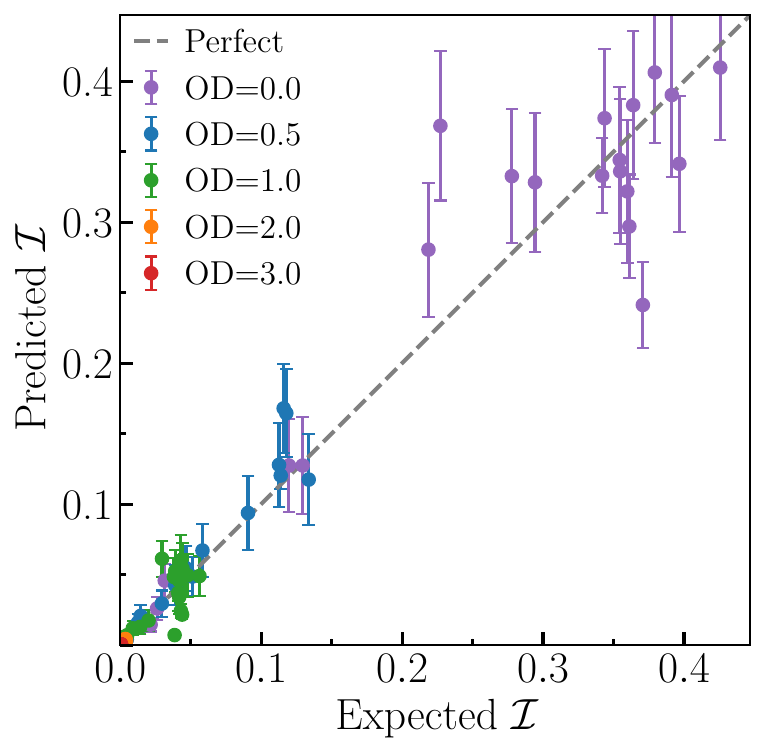}%
\label{fig:parity:irradiance}%
}%
\caption[Parity plots corresponding to greater latent space noise]{Parity plots for \protect\subref{fig:parity:wavelength} wavelength, \protect\subref{fig:parity:log_irradiance} log-normalised irradiance and \protect\subref{fig:parity:irradiance} irradiance obtained using a fixed latent sampling scale of 0.5 instead of the learned standard deviation. Vertical whiskers show the resulting ensemble spread.}
\label{fig:parity}
\end{figure}

\subsection{Model architecture analysis}
\label{sec:robustness:architecture}

To quantify the contribution of the principal architectural and training choices, we perform an ablation analysis relative to the model reported in the main text.
The reference architecture independently decomposes the forward and reverse voltage sweeps using a Legendre basis with maximum degree $D=8$, retains the voltage-range parameters as input features and employs self-supervised pretraining.
Four alternative configurations are considered:

\begin{enumerate}
\item separate forward/reverse decomposition with $D=8$, but without SSL pretraining,
\item separate forward/reverse decomposition with $D=8$, excluding the voltage-range from the input features,
\item separate forward/reverse decomposition with $D=8$, excluding dark current from the input features,
\item separate forward/reverse decomposition with $D=16$, excluding dark current from the input features,
\item combined forward/reverse decomposition with $D=8$, and
\item combined forward/reverse decomposition with $D=16$.
\end{enumerate}

\noindent
All remaining hyperparameters are held fixed.
The $D=16$ architectures are included to compare models with number of trainable parameters closer to that of the reference model (to verify whether change in model performance is related to number of parameters).
Each configuration is trained and evaluated using the same set of random seeds described in \secref{sec:robustness:random_seed}.
Because all model variants achieve high accuracy for log-normalised irradiance ($R^2>0.96$), the ablation analysis focuses on wavelength reconstruction, which is more sensitive to changes in architecture and training strategy.
The metrics presented here are the deterministic $R^2$ metrics, i.e. the ensemble approach to statistical sampling of the latent space is not utilised here.

\begin{table}[htbp]
\centering
\caption[Model architecture ablation analysis]{Ablation analysis of the principal architectural, training, and feature choices. Values report the mean deterministic wavelength $R^2$ across random seeds, with the minimum and maximum values in parentheses. The reference-model statistics are reproduced from \tabref{tab:r2_stats}.}
\label{tab:ablation}
\begin{tabular}{@{}lccc@{}}
\toprule
Omitted feature & Degree & Parameters & $R^2$ (min, max) \\
\midrule
N/A & 8 & 14,799 & 0.9463 (0.9255, 0.9665) \\
\midrule
SSL pretraining & 8 & 14,799 & 0.9357 (0.8967, 0.9538) \\
Voltage min/max & 8 & 14,539 & 0.9436 (0.8731, 0.9620) \\
Dark current & 8 & 12,459 & 0.8473 (0.7820, 0.8909) \\
Dark current & 16 & 14,539 & 0.8647 (0.8250, 0.8971) \\
Separate decomposition & 8 & 12,459 & 0.8825 (0.8340, 0.9078) \\
Separate decomposition & 16 & 14,539 & 0.9221 (0.8834, 0.9465) \\
\bottomrule
\end{tabular}
\end{table}

The reference architecture provides the highest mean wavelength accuracy, with $R^2=0.9463$ and a relatively narrow range across random seeds (0.9255--0.9665).

\noindent
\textbf{Training.}
Removing SSL pretraining reduces the mean to $R^2=0.9357$, demonstrating a measurable benefit from self-supervised initialisation.

\noindent
\textbf{Feature importance.}
Excluding the voltage-range features produces only a small reduction in the mean $R^2$ to $0.9436$, but substantially broadens the range across seeds ($0.8731$--$0.9620$), indicating that retention of the physical voltage scale primarily improves training stability.

The most pronounced degradation arises from omitting the dark current channel from input features, which drops the mean wavelength $R^2$ to $0.8473$ and widens the seed range considerably ($0.7829$--$0.8909$).
This sharp decline in performance highlights the critical role of dark current in wavelength reconstruction.
Increasing the polynomial degree to $D=16$ recovers part of this performance ($R^2=0.8647$) whilst providing a parameter count comparable to the reference model, but remains substantially below the reference architecture that includes dark current as an input feature.
Interestingly, however, this omission does not severely impact irradiance estimation; the mean irradiance $R^2$ remains relatively high at $0.9719$ for $D=8$, with only a modest range expansion ($0.9610$--$0.9827$) relative to the reference model, suggesting that dark current information is more essential for spectral wavelength prediction than for irradiance. Increasing the polynomial degree does not fix the irradiance and actually makes it worse, with an $R^2$ of $0.9605$ and a range of ($0.9431$--$0.9725$).
Overall, this strengthens the argument that irradiance data is not solely encoded in the magnitude of the current-voltage curves, but also in the shape of the curves.

\noindent
\textbf{Architecture importance.}
The largest structural degradation occurs when the forward and reverse sweeps are combined.
Averaging the two branches reduces the mean wavelength $R^2$ to 0.8825 for $D=8$. Increasing the polynomial degree to $D=16$ recovers part of this performance ($R^2=0.9221$) whilst providing a parameter count comparable to the reference model, but remains substantially below the separately decomposed architecture.

The ablation analysis demonstrates that preserving the directional and hysteretic information contained in the bidirectional voltage sweep is the dominant architectural contribution to wavelength reconstruction, while SSL pretraining improves mean predictive performance, inclusion of the voltage range enhances stability across model realisations, and dark current proves indispensable for accurate wavelength modelling.

\clearpage{}
\section*{References}

\bibliographystyle{unsrt}
\bibliography{references}